\documentclass[a4paper,11pt]{article}
\usepackage{jheppub} % for details on the use of the package, please
\usepackage[utf8]{inputenc}
\usepackage{amsfonts}
\usepackage{amsbsy}
\usepackage{url}
\usepackage{enumerate}
\usepackage[justification=raggedright]{caption}
\usepackage{bbm}

\usepackage{graphicx,color}
\usepackage{amssymb,amsmath}
\usepackage{slashed}
\usepackage{hyperref}
\usepackage{braket}
\usepackage{simplewick}
\usepackage[title]{appendix}
\usepackage{caption}
\usepackage{ascmac}
\usepackage{latexsym}
\usepackage{pifont}          
\usepackage{bm}
\usepackage{comment}

\newcommand{\n}{\nonumber}

\newcommand{\mr}[1]{\mathrm{#1}}

\newcommand{\h}[1]{\hspace{#1}}
\newcommand{\f}[2]{\frac{#1}{#2}}

\makeatletter
\@addtoreset{equation}{section}
\makeatother

\begin{document}

\preprint{YGHP-20-02, KUNS-2807}

\title{Dynamics of Nambu monopole in two Higgs doublet models 
-- Cosmological Monopole Collider --}

\author[a,b]{Minoru~Eto,}
\author[c]{Yu~Hamada,}
\author[b]{Masafumi~Kurachi}
\author[b,d]{and Muneto~Nitta} 

\affiliation[a]{Department of Physics, Yamagata University, Kojirakawa-machi 1-4-12, Yamagata, Yamagata 990-8560, Japan}
\affiliation[b]{Research and Education Center for Natural Sciences, Keio University, 4-1-1 Hiyoshi, Yokohama, Kanagawa 223-8521, Japan}
\affiliation[c]{Department of Physics, Kyoto University, Kitashirakawa, Kyoto 606-8502, Japan}

%\affiliation{Research and Education Center for Natural Sciences, Keio University, 4-1-1 Hiyoshi, Yokohama, Kanagawa 223-8521, Japan}
%\affiliation{Theory Center, High Energy Accelerator Research Organization (KEK), Tsukuba, Ibaraki 305-0801, Japan}
%\affiliation{Institute for Cosmic Ray Research (ICRR), The University of Tokyo, Kashiwa, Chiba 277-8582, Japan}

\affiliation[d]{Department of Physics, Keio University, 4-1-1 Hiyoshi, Kanagawa 223-8521, Japan}
%\affiliation{Research and Education Center for Natural Sciences, Keio University, 4-1-1 Hiyoshi, Yokohama, Kanagawa 223-8521, Japan}

\emailAdd{meto(at)sci.kj.yamagata-u.ac.jp}
\emailAdd{yu.hamada(at)gauge.scphys.kyoto-u.ac.jp}
\emailAdd{kurachi(at)keio.jp}
\emailAdd{nitta(at)phys-h.keio.ac.jp}

\abstract{
We study the dynamics of the Nambu monopole in two Higgs doublet models,
which is a magnetic monopole attached by two topological $Z$ strings ($Z$ flux tubes) from two opposite sides.
The monopole is a topologically stable solution of the equation of motions
when the Higgs potential has global $U(1)$ and $\mathbb{Z}_2$ symmetries. 
In this paper, we consider more general cases without the $\mathbb{Z}_2$ symmetry,
and find that it is no longer a static solution
but moves along the $Z$ string being pulled by the heavier string.
After analytically constructing an asymptotic form of the monopole,
we confirm such a motion using the numerical relaxation method.
In addition, we analyze the real time dynamics of the monopole
based on a point-like approximation.
Consequently, if there were long string networks with the monopoles in the early universe,
the monopole accelerates nearly to the speed of light
emitting electromagnetic radiations as a synchrotron accelerator,
and collides to an anti-monopole on the string.
This collision event, which we call the cosmological monopole collider,
can produce much heavier particles than those we can see today, e.g., at the Large Hadron Collider.
}

\maketitle%\thispagestyle{empty}

%%%%%%%%%%%%%%%%% I N T R O D U C T I O N %%%%%%%%%%%%%%%%%%
\section{Introduction}
Topological solitons in field theories play important roles
in various fields of physics such as particle physics, condensed matter physics and cosmology.
Particularly, magnetic monopoles are prominent examples among them,
which were first discovered by 't Hooft and Polyakov \cite{tHooft:1974kcl,Polyakov:1974ek} in the $SO(3)$ Yang-Mills-Higgs model 
as a field theoretical realization of the Dirac hypothesis \cite{Dirac:1931kp}
providing an explanation for the electric charge quantization.
While theoretically they play crucial roles to study non-perturbative aspects of (non-)supersymmetric field theories
\cite{Nambu:1974zg,Seiberg:1994aj,Seiberg:1994rs},
experimentelly such monopoles have never been found in reality, 
except for condensed-matter analogues \cite{Castelnovo:2007qi,Ray:2014sga}.
For instance, such magnetic monopoles are predicted in all grand unified theories (GUTs) 
\cite{Dokos:1979vu,Lazarides:1980va,Callan:1982au,Rubakov:1982fp}, 
and their search have been extensively conducted.
Nevertheless, no GUT monopoles have been found so far.

On the other hand,
a magnetic monopole configuration in the Standard Model (SM) was first considered by Nambu \cite{Nambu:1977ag},
which is called the Nambu monopole.
Since the Nambu monopole is attached by a string 
and consequently is pulled by the tension of the string,
it cannot be a stable soliton.
Nevertheless the Nambu monopoles might be phenemenologically or cosmologically useful; 
they were suggested to produce primordial magnetic fields before their disappearance
\cite{Vachaspati:2001nb,Poltis:2010yu}.
The reason of the absence of stable monopoles in the SM is its trivial topology, that is,
the vacuum manifold is $S^3$ having a trivial second homotopy group $\pi_2$.
In the same way, the vacuum manifold $S^3$ has 
trivial $\pi_0$ for domain walls and $\pi_1$ for cosmic strings.
Non-topological electroweak $Z$ strings (or magnetic $Z$-fluxes)
\cite{Vachaspati:1992fi,Vachaspati:1992jk,Achucarro:1999it,Brandenberger:1992ys,Barriola:1994ez,Eto:2012kb}
have been studied extensively, 
but they were shown to be unstable in the realistic parameter region 
\cite{James:1992zp,James:1992wb}. 
The Nambu monopoles are the end points of these electroweak $Z$ strings \cite{Nambu:1977ag}.

Incidentally, 
while the SM was established by the discoverly of the 125 GeV Higgs boson ($h$) at the Large Hadron Collider (LHC), 
there remain several problems that are unanswered by the SM.
Two Higgs doublet models (2HDMs),
in which one more Higgs doublet is added to the SM, 
are one of the most popular extensions of the SM 
with a potential to solve unsolved problems of the SM 
(for reviews, see, e.g., Refs.~\cite{Gunion:1989we,Branco:2011iw}).
In addition to the 125 GeV Higgs boson ($h$), 
it has four additional scalar degrees of freedom: 
the charged Higgs bosons ($H_{\pm}$), the CP-even Higgs boson ($H$) and the CP-odd Higgs boson ($A$). 
These additional scalars could be directly produced at the LHC, 
though there is no signal so far today, therefore placing lower bounds on masses of those additional scalar bosons.
Those lower bounds highly depend on parameter choices of the 2HDM. %as well as how SM fermions couple to the two doublets.
For more detailed phenomenological studies, see, e.g., Refs.~\cite{Trodden:1998ym,Kanemura:2015mxa,Kanemura:2014bqa,Kling:2016opi,Haller:2018nnx} and references therein.
One of the most remarkable aspects of 2HDMs  
distinguishable from the SM may be
that it has a much richer vacuum structure than the SM,
thereby allowing a variety of 
topologically stable solitons, 
in addition to non-topological solitons \cite{La:1993je,Earnshaw:1993yu,Perivolaropoulos:1993gg,Bimonte:1994qh,Ivanov:2007de,Brihaye:2004tz,Grant:2001at,Grant:1998ci,Bachas:1996ap} analogous to the SM;
domain walls \cite{Battye:2011jj,Brawn:2011,Eto:2018tnk,Eto:2018hhg,Chen:2020soj,Battye:2020sxy}, 
membranes \cite{Bachas:1995ip,Riotto:1997dk}, 
and cosmic strings 
such as topological $Z$ strings \cite{Dvali:1993sg,Dvali:1994qf,Eto:2018tnk,Eto:2018hhg}
(see also Ref.~\cite{Bachas:1998bf}). 
However, magnetic monopoles were not examined because of the trivial second homotopy group $\pi_2$ of the 2HDM, as in the SM.

In the previous paper \cite{Eto:2019hhf}, the present authors studied the Nambu monopole in the 2HDM, 
which is a magnetic monopole attached by {\it two topological $Z$ strings from two opposite sides}.
The monopole is topologically stable and a regular solution of the equation of motion (EOM)
when the Higgs potential has two global symmetries;
One is a global $U(1)$ symmetry that ensures the stability of the topological $Z$ strings.
The other is a discrete symmetry $\mathbb{Z}_2$ exchanging the topological $Z$ strings.
%The stability of the monopole is topologically protected by a combination of the two symmetries.
The string tensions pulling the monopole are balanced due to the $\mathbb{Z}_2$ symmetry,
and thus the monopole does not move unlike the Nambu monopole in the SM and can be regarded as a topologically stable $\mathbb{Z}_2$ kink on one string.

If the symmetries were realized in nature, the monopoles are expected to be abundant in the early and present universe.
However, the models with these two symmetries are phenomenologically disfavored.
The $U(1)$ symmetry should be explicitly broken to give the mass to the CP-odd Higgs boson
and the $\mathbb{Z}_2$ symmetry is broken by the Yukawa coulings with the SM fermions.
Therefore, these two symmetries may not be realized in the Higgs potential, at least for phenomenologically viable 2HDMs,
and the monopoles may not be exactly stable.
In such a case, it is still important to investigate whether (un)stable 
monoples can exist, and if so, how (un)stable they are.
If they are sufficiently stable even for the non-symmetric case, 
they can be abundant and might be observed in the current monopole searches.
In addition, they could be useful to impose constraints on the parameter space of 2HDMs
from cosmological arguments such as the cosmological monopole problem.

In this paper, we investigate the dynamics of the Nambu monopole for the non-symmetric case.
Especially, we consider the case without the $\mathbb{Z}_2$ symmetry,
in which the tensions of the $Z$ strings are not degenerated.
As a result, the monopole is no longer static, but moves being pulled by the heavier string.
We can confirm this picture based on the numerical relaxation.
Furthermore, the monopole continues to accelerate emitting electromagnetic radiations.
If the monopoles and networks of the $Z$ strings are produced in the early universe,
a monopole on a long string accelerates sufficiently and reaches nearly to the speed of light.
This can be regarded as an accelerating charged particle 
in a synchrotron accelerator with a radius of the horizon scale.
After the acceleration,
the monopole eventually collides to an anti-monopole on the string
and would produce various high energy radiations and particles.
The typical kinetic energy of the accelerated monopole is
$\sim (\cos 2\beta)^{1/4}~ 10^8 ~\mathrm{TeV}$
with $\tan \beta$ being the ratio of the vacuum expectation values of the two Higgs doublets.
This is a quite high-energy event in the early universe, which we call as the Cosmological Monopole Collider.
Consequently, it is unlikely that the monopoles without the $\mathbb{Z}_2$ symmetry remain abundant in the present universe.
Instead, some remnants of the collision events could be detected by astrophysical and cosmological observations.

This paper is organized as follows.
Sec. \ref{172002_17Nov19} is devoted to introduce the model.
We will explain the $U(1)$ symmetry and the $\mathbb{Z}_2$ symmetry
and show mass spectrum in the model.
In Sec.~\ref{172423_16Jan20}, the electroweak strings will be considered.
We will first review the topological $Z$ strings, and then analyze string tensions of the general electroweak strings.
In addition, we will investigate asymptotic properties of the $Z$ strings.
In Sec.~\ref{202730_25Feb20}, we will consider the Nambu monopole in the 2HDM without the $\mathbb{Z}_2$ symmetry.
We will construct an asymptotic form of the monopole
and observe that the magnetic flux spherically spreads from the monopole independently of $\tan\beta$.
After that, we will give a cylindrical ansatz describing the monopole
and perform the numerical relaxation.
We will show results of the relaxation for several parameter choices,
in which the monopole slowly moves being pulled by the heavier string.
In Sec.~\ref{105733_21Nov19} we will analyze the real dynamics of the Nambu monopole
based on an approximation where we regard the monopole as a point particle.
Consequently, we find that it accelerates immediately nearly to the speed of light by the CMC.
Finally we will give a discussion and conclusion in Sec.~\ref{122632_28Nov19}.
In Appenix \ref{165251_19Nov19}, we will provide a derivation for expressions of the gauge fields
in the presence of the monopole configuration of the Higgs field.

%%%%%%%%%%%%%%%%%%%%%%%%%%%%%%%%%%%%%%%%%%%%%%%%%%%%
\section{The model}\label{172002_17Nov19}
\subsection{The Lagrangian and Higgs potential}
We introduce two $SU(2)$ doublets, $\Phi_1$ and $\Phi_2$, both with the hypercharge $Y=1$.
The Lagrangian which describes the electroweak and Higgs sectors is written as
\begin{align}
\h{-1em} {\mathcal L} = - \frac{1}{4}\left(Y_{\mu\nu}\right)^2 - \frac{1}{4}\left(W_{\mu\nu}^a\right)^2  + \left|D_\mu \Phi_i \right|^2 - V(\Phi_1, \Phi_2).
\label{eq:L}
\end{align}
Here, $Y_{\mu\nu}$ and $W^a_{\mu\nu}$ describe field strength tensors of the hypercharge
and weak gauge interactions, respectively, with $\mu$ ($\nu$) and $a$ being Lorentz and weak iso-spin indices, respectively. 
$D_\mu$ represents the covariant derivative acting on the Higgs fields, and the index $i$ runs $i=1,2$.
The most generic quartic potential $V(\Phi_1, \Phi_2)$ for the two Higgs doublets is given by
\begin{align}
V(\Phi_1,\Phi_2) & = m_{11}^2\Phi_1^\dagger\Phi_1 + m_{22}^2 \Phi_2^\dagger \Phi_2 
 - \left(m_{12}^2 \Phi_1^\dagger \Phi_2 + {\rm h.c.}\right) \n\label{173450_17Nov19} \\
 & + \frac{\beta_1}{2}\left(\Phi_1^\dagger\Phi_1\right)^2 + \frac{\beta_2}{2}\left(\Phi_2^\dagger\Phi_2\right)^2 
  + \beta_3\left(\Phi_1^\dagger\Phi_1\right)\left(\Phi_2^\dagger\Phi_2\right) 
   + \beta_4 \left(\Phi_1^\dagger\Phi_2\right)\left(\Phi_2^\dagger\Phi_1\right)\n \\
 &+ \left\{\frac{\beta_5}{2}\left(\Phi_1^\dagger\Phi_2\right)^2 + {\rm h.c.}\right\},
\end{align}
where we have imposed a (softly-broken) ${\mathbb Z}_2$ symmetry,
$\Phi_1 \to +\Phi_1$,  $\Phi_2 \to -\Phi_2$,
in order to suppress Higgs-mediated flavor-changing neutral current processes.
In this paper, we assume that both the Higgs fields develop real vacuum expectation values (VEVs)
as $ \Phi_1 = \left( 0,  v_1\right) ^T, \Phi_2 =  \left(0 , v_2\right)^T $.\footnote{
Note missing ``$\sqrt{2}$'' in our notation for the VEVs.}
Then the electroweak scale, $v_\mr{EW}$ ($\simeq $ 246 GeV), can be expressed by these VEVs as $v_{\rm EW}^2 = 2 v_{\rm sum}^2 \equiv 2 (v_1^2 + v_2^2)$.

For later use, we rewrite the Higgs fields in a two-by-two matrix form\cite{Grzadkowski:2010dj}, $H$,
defined by 
\begin{equation}
 H = \left( i\sigma_2 \Phi_1^*,\ \Phi_2\right).
\end{equation}
The matrix field $H$ transforms under the electroweak $SU(2)_L \times U(1)_Y$ symmetry as
\begin{equation}
H \to \exp\left[\f{i}{2}\theta_a(x) \sigma_a\right] H ~\exp\left[-\f{i}{2} \theta_Y(x) \sigma_3\right],
\end{equation}
where the group element acting from the left belongs to $SU(2)_L$ and the other element acting from the right beongs to $U(1)_Y$.
Therefore the covariant derivative on $H$ can be expressed as
\begin{equation}
D_\mu H =\partial_\mu H - i \frac{g}{2} \sigma_a W_\mu^a H + i \frac{g'}{2}H\sigma_3 Y_\mu.
\end{equation}
The VEV of $H$ is expressed by a diagonal matrix $\langle H \rangle = \mr{diag} (v_1,v_2)$,
and the Higgs potential can be written by using $H$ as follows:
\begin{align}
V(H)
& = - m_{1}^2~ \mr{Tr}|H|^2 - m_{2}^2~ \mr{Tr}\left(|H|^2 \sigma_3\right) - \left( m_{3}^2 \det H + \mr{h.c.}\right)\nonumber\label{173547_17Nov19} \\
& + \alpha_1~\mr{Tr}|H|^4  +  \alpha_2 ~\left(\mr{Tr}|H|^2 \right)^2+ \alpha_3~ \mr{ Tr}\left(|H|^2 \sigma_3 |H|^2\sigma_3\right)  \n \\
& + \alpha_4~ \mr{Tr}\left(|H|^2 \sigma_3 |H|^2\right)+ \left(\alpha_5 \det H^2 + \mr{h.c.}\right),
\end{align}
where $|H|^2 \equiv H ^\dagger H$
and the relations between the parameters in Eq.~(\ref{173547_17Nov19}) and in Eq.~(\ref{173450_17Nov19}) are given by 
\begin{align}
  m_{11}^2 = -m_1^2 - m_2^2 , \h{2em}  m_{22}^2 = -m_1^2 + m_2^2 ,  \h{2em}  m_{12} = m_3 ,\\
  \beta_1  =2(\alpha_1+ \alpha_2 + \alpha_3 +  \alpha_4 ), \h{2em} \beta_2  =2( \alpha_1+ \alpha_2 + \alpha_3 -  \alpha_4 ), \\
  \beta_3 = 2(\alpha_1+ \alpha_2 - \alpha_3) ,  \h{2em} \beta_4 = 2(\alpha_3 - \alpha_1),  \h{2em} \beta_5 = 2\alpha_5.
\end{align}

\subsection{$U(1)_a$ symmetry}

In the rest of this paper we will mostly restrict ourselves to the limited parameter space 
\begin{equation}
\text{$U(1)_a$ condition}:\quad m_3 = \alpha_5 = 0.
\label{eq:cond_1}
\end{equation}
When %$m_3=\alpha_5=0$ in Eq.~(\ref{173547_17Nov19}), 
this condition holds, 
the Lagrangian is invariant under a global $U(1)_a$ transformation,
which is defined by a rotation of the relative phase of the two doublets:
$ H \to e ^ {i \alpha} H$ (or $\Phi_1 \to e^{-i\alpha} \Phi_1$, $\Phi_2 \to e^{i\alpha} \Phi_2$) ($0\leq\alpha < 2\pi$).
After $H$ gets the VEV, this $U(1)_a$ symmetry is spontaneously broken
and the corresponding Nambu-Goldstone (NG) boson appears,
which is nothing but the CP odd Higgs boson ($A$). 
The spontaneously broken $U(1)_a$ symmetry gives rise to non-trivial topological excitations in which we are interested
in this work.

Because an experimental lower bound on the mass of $A$ is typically $\mathcal{O}(100)$ GeV
(which highly depends on how the doublets couple to the SM fermions),
such a massless $A$ is phenomenologically disfavored.
Therefore, in realistic cases,
we should break the $U(1)_a$ symmetry explicitly by switching on $m_3$ and $\alpha_5$, giving a mass to $A$.
Nevertheless, we set $m_3=\alpha_5=0$ throughout this paper
because otherwise an analysis of the dynamics of the monopole becomes very complicated as we see in Sec.~\ref{122632_28Nov19}.

\subsection{Custodial symmetry and $(\mathbb{Z}_2)_{\rm C}$ symmetry}
The custodial transformation acting on the matrix $H$ \cite{Grzadkowski:2010dj,Pomarol:1993mu}
is defined as the following global $SU(2)$ transformation: $H \to U H U ^\dagger$, $U \in SU(2)_\mr{C}$.\footnote{
Note that this  $SU(2)_\mr{C}$ transformation is different from the $U(2)$ basis transformation:
 $\Phi_i \to \sum_{j=1}^2M_{ij} \Phi_j$, $M \in U(2)$ ($i=1,2$).
 }
In addition, the $SU(2)_W$ gauge field transforms as an adjoint representation simultaneously.
The potential $V(H)$ given in Eq.~(\ref{173547_17Nov19}) is invariant under this transformation,
which we call as the custodial symmetry, when $m_2=\alpha_3=\alpha_4=0$.
Note that the gauge kinetic term of $H$ cannot be invariant under this transformation
because of the presence of the $U(1)_Y$ gauge field.
Thus the custodial symmetry is not exact symmetry of the theory 
but is explicitly broken by the gauge interaction.

The bosonic Lagrangian ${\cal L}$ in Eq.~(\ref{eq:L}) 
still has a symmetry under a $\mathbb{Z}_2$ transformation defined by
\begin{align}
 \begin{cases}
  H \to i\sigma_1 H (i \sigma_1)^\dagger \\
  W_\mu \to  i\sigma_1 W_\mu (i \sigma_1)^\dagger \\
  Y_\mu \to - Y_\mu
 \end{cases}
\end{align}
which we call as the $(\mathbb{Z}_2)_\mr{C}$ symmetry.
Since $i \sigma_1 \in SU(2)_\mr{C}$,
$(\mathbb{Z}_2)_\mr{C}$ acts on $H$ and $W_\mu$ as a subgroup of the $SU(2)_\mr{C}$ transformation, but not for $Y_\mu$.
Imposing the $(\mathbb{Z}_2)_\mr{C}$ symmetry on ${\cal L}$ in Eq.~(\ref{eq:L}) reads
\begin{equation}
\text{$(\mathbb{Z}_2)_\mr{C}$ condition}:\quad m_2 = \alpha_4 = 0.
\label{eq:cond_2}
\end{equation}
Note that,
as we will manifestly show below,  
the $(\mathbb{Z}_2)_\mr{C}$ is not spontaneously broken in the vacuum because of 
$\left<H\right> \propto {\bf 1}_2$ ($\tan \beta \equiv v_2/v_1 =1$).
Since the transformation of $H$ can be written as
$\Phi_1 \to i \sigma_2 \Phi_2^\ast $ and $ \Phi_2 \to i \sigma_2 \Phi_1^\ast $,
 it can be regarded as a combination of exchanging the two doublets and the CP transformation in the Higgs sector.

Similarly to the $U(1)_a$ symmetry,
 the $(\mathbb{Z}_2)_\mr{C}$ symmetry may not be realized in Lagrangian for realistic models
because it is broken by the Yukawa couplings between the doublets and SM fermions.
For instance, in the type-I 2HDM,
the Yukawa couplings are given by 
\begin{equation}
 \mathcal{L}_{\rm yukawa}=  y_e \bar{e}_R \Phi_2^\dagger L_L + y_u \bar{u}_R \tilde{\Phi}_2^\dagger Q_L + y_d \bar{d}_R \Phi_2^\dagger Q_L ,\h{2em} (\tilde{\Phi}_2\equiv i \sigma_2 \Phi_2^\ast)
\end{equation}
which is clearly not invariant under the $(\mathbb{Z}_2)_\mr{C}$ transformation.
As a result, $m_2$ and $\alpha_4$ should be generated by radiative corrections from the SM fermion loops 
even when the $(\mathbb{Z}_2)_\mr{C}$ condition in Eq.~(\ref{eq:cond_2}) is imposed at the tree level.

We studied the Nambu monopole in the 2HDM under both the $U(1)_a$ and $(\mathbb{Z}_2)_\mr{C}$ conditions
in our previous study \cite{Eto:2019hhf}.
In this paper, we relax the latter and study the Nambu monopoles in more realistic cases.
Namely, we will investigate the Nambu monopoles in the cases without the $(\mathbb{Z}_2)_{\rm C}$ condition by setting
$m_2 \neq 0,~ \alpha_4 \neq 0,~ \alpha_3 \neq 0$ in the Higgs potential.

\subsection{Higgs mass spectrum}

For the $U(1)_a$ symmetric Higgs potential with the condition (\ref{eq:cond_1}), the Higgs VEVs are given by
\begin{equation}
 v_1 =\frac{\sqrt{(\alpha_1+\alpha_3-\alpha_4) m_1^2 + (\alpha_1+2\alpha_2 + \alpha_3-\alpha_4)m_2^2}}{\sqrt{2(\alpha_1 + \alpha_3)(\alpha_1+2\alpha_2+\alpha_3)-2\alpha_4^2}},
\end{equation}
\begin{equation}
v_2 = \frac{\sqrt{(\alpha_1+\alpha_3+\alpha_4)m_1^2 - (\alpha_1+2\alpha_2 + \alpha_3+\alpha_4)m_2^2}}{\sqrt{2(\alpha_1 + \alpha_3)(\alpha_1+2\alpha_2+\alpha_3)-2\alpha_4^2}},
\end{equation}
and $\tan \beta$ is defined as
\begin{equation}
 \tan \beta \equiv \f{v_2}{v_1}
  = \f{\sqrt{(\alpha_1+\alpha_3+\alpha_4)m_1^2 - (\alpha_1+2\alpha_2 + \alpha_3+\alpha_4)m_2^2}}
  {\sqrt{(\alpha_1+\alpha_3-\alpha_4) m_1^2 + (\alpha_1+2\alpha_2 + \alpha_3-\alpha_4)m_2^2}}.
\end{equation}
Now, it is clear that $\tan\beta = 1$ holds if we impose the $(\mathbb{Z}_2)_\mr{C}$ condition.

In the matrix notation, fluctuations around the VEVs can be parametrized as
\begin{align}
 H &= \mr{diag}(v_1,v_2) + \f{1}{\sqrt{2}}\left(\chi^A + i \pi^A \right) \sigma^A ~ \mr{diag}(1,\tan \beta) \h{2em} (A=0,\cdots,3)\label{010632_18Mar20}
\end{align}
with $\sigma^A = (\bm{1},\sigma^a)$ ($a=1,2,3$).
Here $\pi^0$ is the NG boson for the $U(1)_a$ symmetry,
which is called as the CP-odd Higgs boson $A$ in the literature.
On the other hand, $\pi^a$'s  are would-be NG bosons for $SU(2)_W\times U(1)_Y$ and eaten by the gauge bosons.

When the potential has the $(\mathbb{Z}_2)_\mr{C}$ symmetry, yielding $\tan\beta=1$,
$(\chi^1 \pm i\chi^2)/\sqrt{2}$ are called as the charged Higgs bosons $H^\pm$, respectively.
On the other hand, $\chi^0$ and $\chi^3$ are CP-even neutral components.
Note that they are not mixed
because they are $(\mathbb{Z}_2)_\mr{C}$ even and odd, respectively.
The lighter one is identified with the SM Higgs boson $h$
while the other is called as the heavier CP-even neutral Higgs boson $H$.
Their masses are given by
\begin{align}
 (m_{H^\pm})^2 = \f{2(\alpha_1-\alpha_3)m_1^2}{\alpha_1 + 2 \alpha_2 + \alpha_3} , \h{2em} (m_{\chi^0})^2=2m_1^2, \h{2em}
 (m_{\chi^3})^2 = \f{2(\alpha_1+\alpha_3)m_1^2}{\alpha_1 + 2 \alpha_2 + \alpha_3} .
\end{align}
The lighter one among $\chi^0$ and $\chi^3$ is identified with the SM Higgs boson $h$
while the other is called as the heavier CP-even neutral Higgs boson $H$.

On the other hand, it is more complicated for the potential without the $(\mathbb{Z}_2)_\mr{C}$ symmetry.
The charged Higgs bosons are given by more complicated linear combinations of $\chi^1$ and $\chi^2$.
In addition, $\chi^3$ and $\chi^0$ are mixed with the mixing angle $\alpha$,
which is defined by
\begin{equation}
 \tan2\alpha = \f{2 B_S}{A_S-C_S},
\end{equation}
where
\begin{equation}
 A_S\equiv \f{2 \left(\alpha_{134}^{(+)}  + \alpha_2 \right) \left(m_1^2 \alpha_{134}^{(-)} + 
  m_2^2 (\alpha_{134}^{(-)} + 2 \alpha_2 )\right)}
  { (\alpha_1 + \alpha_3) (\alpha_1 + 2 \alpha_2 + \alpha_3) -  \alpha_4^2},
\end{equation}
\begin{equation}
 B_S\equiv
  \f{2 \alpha_2 \sqrt{ m_1^2 \alpha_{134}^{(-)} +  m_2^2 \left(\alpha_{134}^{(-)} + 2 \alpha_2 \right)}
 \sqrt{ m_1^2 \alpha_{134}^{(+)} - m_2^2 \left(\alpha_{134}^{(+)} + 2 \alpha_2 \right)}}
  {(\alpha_1 + \alpha_3) (\alpha_1 + 2 \alpha_2 + \alpha_3) - \alpha_4^2},
\end{equation}
\begin{equation}
C_S\equiv \f{2 \left(\alpha_{134}^{(-)} + \alpha_2 \right) \left(m_1^2 \alpha_{134}^{(+)}
  - m_2^2 (\alpha_{134}^{(+)} + 2 \alpha_2 ) \right)}
{(\alpha_1 + \alpha_3) (\alpha_1 + 2 \alpha_2 + \alpha_3) - \alpha_4^2},
\end{equation}
with $\alpha_{134}^{(\pm)} \equiv \alpha_1+\alpha_3 \pm \alpha_4$.
After the rotation by $\alpha$, $\chi^0$ and $\chi^3$ are transformed into the mass eigen states $H$ and $h$.
We regard the lighter one as the SM Higgs again.
The mass eigen values for the Higgs bosons are given by
\begin{align}
 m_H^2 &= 2 \left[v_{\rm sum}^2\alpha_{123} +(v_1^2-v_2^2)\alpha_4 + \sqrt{\left( (v_1^2-v_2^2)\alpha_{123}+ v_{\rm sum}^2 \alpha_4\right)^2 + 4 \alpha_2^2v_1^2v_2^2}\, \right],\\
 m_h^2 &= 2 \left[v_{\rm sum}^2\alpha_{123}+(v_1^2-v_2^2)\alpha_4- \sqrt{\left( (v_1^2-v_2^2)\alpha_{123} + v_{\rm sum}^2 \alpha_4\right)^2 + 4 \alpha_2^2v_1^2v_2^2}\, \right],\\
  m_{H^\pm}^2 &= 2 (\alpha_1-\alpha_3)v_{\rm sum}^2,\label{112349_16Jan20}
\end{align}
where $\alpha_{123}\equiv \alpha_1+\alpha_2+\alpha_3$ and $v^2_{\rm sum} = v_1^2 + v_2^2$.

The mass of the gauge bosons are given by
\begin{equation}
m_W^2 = \frac{g^2v_{\rm sum}^2}{2} = \frac{g^2v_{\rm EW}^2}{4},\quad
m_Z^2 = \frac{g^2v_{\rm sum}^2}{2\cos^2\theta_\mr{W}} = \frac{g^2v_{\rm EW}^2}{4\cos^2\theta_\mr{W}},
\end{equation}
with the standard definitions $\cos\theta_\mr{W} = \frac{g}{\sqrt{g^2 + g'{}^2}}$,
$Z_\mu \equiv W_\mu ^3 \cos\theta_\mr{W} - Y_\mu \sin\theta_\mr{W}$, and $A_\mu\equiv W_\mu ^3 \sin\theta_\mr{W} + Y_\mu \cos\theta_\mr{W}$.

%

%

%%%%%%%%%%%%%%%%%%%%%%%%%%%%%%%%%%%%%%%%%%%%%%%%%%%%
\section{Electroweak strings}\label{172423_16Jan20}
In Refs.\cite{Eto:2018tnk,Eto:2018hhg,Dvali:1994qf,Dvali:1993sg}, it is pointed out that, unlike in the SM case, 2HDMs allow topologically stable strings to exist thanks to the global $U (1)_a$ symmetry.
First, consider topological strings with the $Z$ flux (topological $Z$ strings).
There are two types of topological $Z$ strings corresponding to which one of the two Higgs doublets is to be wound.
To see that, let us take $W_\mu^\pm = A_\mu=0$.

\subsection{$Z$ strings}
The solution called the $(1,0)$-string on the $z$ axis 
%\footnote{$(1,0)$ means that the phase of $\Phi_1$ winds once around the circle at spatial infinity but that of $\Phi_2$ does not.}
is given by
\begin{align}
 H^{(1,0)} &= v_{\rm sum}\begin{pmatrix}
	        f^{(1,0)}(\rho) e^{i\varphi} \cos \beta&0 \\
               0 & h ^{(1,0)} (\rho) \sin\beta  
	      \end{pmatrix},\label{eq:H(1,0)}\\
 Z_i ^{(1,0)} &=  -\cos^2 \beta \f{2  \cos \theta_\mr{W}}{g}
 \f{ \epsilon_{3ij}x^j}{\rho^2} \left(1- w ^{(1,0)}(\rho)\right)\label{013304_3Dec19},
\end{align}
where $\rho \equiv \sqrt{x^2+y^2}$ and $\varphi$ is the rotation angle around the $z$-axis. 
The boundary conditions imposed on the profile functions are $f ^{(1,0)} (0)={h^{(1,0)}}'(0)=w ^{(1,0)} (\infty)=0$, $w^{(1,0)} (0)=f^{(1,0)}(\infty)=h^{(1,0)}(\infty)=1$.
Thus, the asymptotic form of $H^{(1,0)}$ at $\rho \to \infty$ 
is $\sim v_{\rm sum} \exp[{\f{i\varphi}{2}}] ~\mr{diag} \left(\cos\beta, \sin\beta \right)\exp[{\f{i\varphi}{2} \sigma_3}] $.

On the other hand, the solution called the $ (0,1) $-string is given by 
\begin{align}
 H^{(0,1)} &= v_{\rm sum} \begin{pmatrix}
	        h^{(0,1)} (\rho)\cos\beta &0 \\
               0 & f ^{(0,1)} (\rho) e^{i\varphi} \sin\beta
	      \end{pmatrix},\\
 Z_i ^{(0,1)} &= \sin^2 \beta \f{2  \cos \theta_\mr{W}}{g}  \f{ \epsilon_{3ij}x^j}{\rho^2} \left(1- w ^{(0,1)}(\rho)\right)\label{012222_17Mar19},
\end{align}
and $H^{(0,1)} \to v_{\rm sum} \exp[{\f{i\varphi}{2}}] ~\mr{diag} \left(\cos\beta, \sin\beta \right)\exp[{\f{-i\varphi}{2} \sigma_3}] $ for $\rho \to \infty$.
The boundary conditions for $f ^{(0,1)}$, $h^{(0,1)}$ and $ w ^{(0,1)}$ are the same as the $(1,0)$-string.

Looking at the asymptotic forms, it is clear that 
both the $(1,0)$- and $(0,1)$-strings have winding number $1/2$ for the global $U(1)_a$ symmetry,
and thus they are topological vortex strings of the global type. Similarly to standard global vortices, their tensions 
(masses per unit length)
logarithmically diverge. It can be seen
%Note that both of them have the same logarithmically divergent tension due to 
from the kinetic term of the Higgs field:
\begin{equation}
 2\pi \int d\rho \rho ~\mr{tr}|D_i H^{(1,0)}|^2
  \sim 2\pi \int d\rho \rho ~\mr{tr}|D_i H^{(0,1)}|^2
  \sim 2\pi \sin ^2 \beta \cos ^2 \beta ~v_\mr{sum}^2 \int \f{d\rho} {\rho} \label{202812_20Dec19}
\end{equation}
for $\rho\to \infty$.
Note that $\beta$ dependence of $Z_i$ in Eqs.~(\ref{013304_3Dec19}) and (\ref{012222_17Mar19}) were determined in such a way
that the logarithmic divergences are minimized.
For $\tan \beta=1$, it becomes a quarter of that for a global $U(1)_a$ {\it integer vortex}
because of the half winding number for $U(1)_a$ \cite{Eto:2018tnk}.

On the other hand,
they also have a winding number $1/2$ inside the gauge orbit $U(1)_Z\in SU(2)_W\times U(1)_Y$,
which lead to the $Z$ fluxes flowing inside them.
The amounts of the fluxes of $(1,0)$- and $(0,1)$-string are
\begin{equation}
 \Phi_Z^{(1,0)}= \cos ^2 \beta \frac{4 \pi \cos \theta_\mr{W} }{ g}, \h{2em}  \Phi_Z^{(0,1)}= -\sin^2 \beta \frac{4 \pi  \cos \theta_\mr{W} }{ g}, 
\label{eq:Z_fluxes}
\end{equation}
along the $z$-axis, respectively.
For $\tan \beta =1$, they are half of that of a non-topological $Z$ string in the SM because of the half winding number.
%and their ratio is $\tan^2\beta$.
Note that the $Z$ fluxes for the $(1,0)$- and $(0,1)$-strings are different for generic $\beta$, 
unlikely the logarithmic divergent energy. 
The $Z$ flux is squeezed into a flux tube.
It decays exponentially fast as a usual Abrikosov-Nielsen-Olesen vortex 
\cite{Abrikosov:1956sx,Nielsen:1973cs} in the Abelian-Higgs model, in contrast to the $1/\rho$ tail given in Eq.~(\ref{202812_20Dec19}).
In other words, contributions to the energy from the non-Abelian parts do not diverge. 
Therefore, the difference of the tensions of the $(1,0)$- and $(0,1)$- strings appears only in finite portion of the tensions,
which are due to the non-symmetric $Z$ fluxes and the Higgs potential energy for $\tan\beta\neq 1$.

It is instructive to see the $Z$ strings from the special point $m_2 = \alpha_3 = \alpha_4 =0$.
As we explained above, the Higgs potential has the custodial $SU (2)_\mr{C} $ symmetry there. Furthermore,
if $\sin \theta_\mr{W}=0 $, the custodial symmetry is an exact symmetry of the Lagrangian.
However, the presence of a topological string solution spontaneously breaks it down to a $ U(1)_\mr{C}$ subgroup in the vortex core,
giving $S^2$ ($\simeq SU(2)_\mr{C}/ U(1)_\mr{C}$) orientational moduli to the vortex,
as studied in Refs.\cite{Eto:2018tnk,Eto:2018hhg}.
Each point on the $S^2$ moduli space corresponds to a physically different string solution with a different magnetic flux, 
having a common winding number $1/2$ for the global $U(1)_a$.
We parametrize the $S^2$ moduli space by two parameters $0\leq \zeta \leq \pi$, $0\leq \chi<2\pi$,
where $\zeta$ and $\chi$ correspond to the zenith and azimuth angles, respectively.
We identify the (1,0)-string, Eq.~(\ref{012222_17Mar19}),
as the one associated with the north pole of the $S^2$ moduli space, $\zeta = \pi$.
On the other hand, the (0,1)-string, Eq.~(\ref{013304_3Dec19}), corresponds to the south pole, $\zeta = 0$.
A string solution on a generic point of the $S^2$ moduli space can be obtained
by acting an $SU(2)_\mr{C}$ transformation on the $(1,0)$-string.

However, $ \sin \theta_\mr{W} \neq 0 $ in nature, and it explicitly breaks the custodial symmetry even when $m_2=\alpha_3=\alpha_4=0$.
As a consequence, almost all the points of the $S^2$ moduli space are energetically lowered, except for the equator $\zeta = \pi/2$.
As studied in Refs.\cite{Eto:2018tnk,Eto:2018hhg}, the two $Z$ strings, $(1,0)$-string and $(0,1)$-string, 
are the most stable with degeneracy among the topological strings.
On the other hand, the strings corresponding to the equatorial points of $S^2$,
which contain a $W$ flux and are called as $W$ strings, are the most unstable.
The effect of $\alpha_3 \neq 0$ also breaks the custodial symmetry
and lifts or lowers the tensions on the moduli space depending on the sign of $\alpha_3$.
Note that it does not break the $(\mathbb{Z}_2)_\mr{C}$ symmetry because the $(\mathbb{Z}_2)_{\rm C}$ condition is still satisfied,
so that it keeps the degeneracy of the two $Z$ strings as we will see in the next subsection.
However, $m_2 \neq 0$ and $\alpha_4 \neq 0$ break the degeneracy.%
\footnote{The symmetry under rotations around the $ \sigma^3 $-axis still remains as $ U(1)_\mr{EM} $.}
They make the energetic structure of the ``moduli space'' quite complicated.
In the next subsection, we investigate the effects of $m_2 \neq 0$, $\alpha_3 \neq 0$ and $\alpha_4 \neq 0$
on the string tensions.

\subsection{Tensions for the electroweak strings}\label{152151_15Jan20}
Let us investigate the effects of $m_2 \neq 0$, $\alpha_3 \neq 0$ and $\alpha_4 \neq 0$ on the string tensions.
In Ref.~\cite{Eto:2018tnk}, it was numerically studied for various parameter regions.
%for the $(\mathbb{Z}_2)_\mr{C}$ symmetric cases.
In this subsection, we analytically study it both for the case with and without the $(\mathbb{Z}_2)_\mr{C}$ symmetry.
For simplicity, we consider $\theta_\mr{W}=0$.%
\footnote{The effect of $\theta_\mr{W} \neq 0$ is only lowering the tensions of the $Z$ strings slightly.}

We begin with the string for $m_2^2 = \alpha_3 = \alpha_4=0$.
Because of the custodial symmetry, the string solution has the moduli parameters which are the coordinates 
$\zeta$ ($0 \leq \zeta \leq \pi$) and $\chi$ ($0 \leq \chi \leq 2\pi$) of the moduli space $S^2$.
Since the azimuth angle $\chi$ remains as the moduli even when we turn on $m_2$, $\alpha_3$, and $\alpha_4$,
we fix it as $\chi = 0$. Then the generic solution for $\zeta$ is given by
\begin{eqnarray}
 H(\zeta) &=& U  \begin{pmatrix}
		v h_0(\rho)&0 \\ 0& v e^{i \varphi }f_0(\rho)
  \end{pmatrix} U^\dagger,\\
 W_i(\zeta) &=&  -\f{1}{g}  \f{ \epsilon_{3ij}x^j}{\rho^2} \left(1- w_0(\rho)\right) U \f{\sigma_3}{2} U^\dagger,
\end{eqnarray}
where we have ntroduced
\begin{equation}
 U \equiv \exp\left[i \f{\zeta}{2} \sigma_2\right] , \h{2em} v\equiv \f{m_1}{\sqrt{2 \alpha_1 + 4 \alpha_2}}.
\end{equation}
Here $f_0(\rho)$, $h_0(\rho)$ and $w_0(\rho)$ are determined by the EOMs for $m_2^2, \alpha_3,\alpha_4=0$.
The $(0,1)$- and $(1,0)$-string correspond to $\zeta = 0$ and $\pi$, respectively.
The string tensions are degenerated for all $\zeta$ as long as the condition  $m_2^2 = \alpha_3 = \alpha_4=0$ is kept.

Next, we turn on the $(\mathbb{Z}_2)_\mr{C}$ breaking parameters, {\it i.e.},
$m_2^2$, $\alpha_3$, and $\alpha_4$, and estimate their effects on the string tensions.
We use a perturbation with respect to the parameters assuming they are sufficiently small.
Thus, the tension of the perturbed string is approximated
by substituting the unperturbed string solution $\{H(\zeta), W_i(\zeta)\}$
into the energy functional.
Then, we express the tension as follows:
\begin{align}
 T(\zeta) &= T_0 + \Delta T(\zeta),
\end{align}
where $T_0$ is the $\zeta$-independent part
and $\Delta T$ is the $\zeta$-dependent one, which is caused by the breaking of the custodial symmetry.
After some algebra, $\Delta T(\zeta)$ is obtained as
\begin{eqnarray}
\label{031720_1Feb20}
\Delta T(\zeta) = A_0 + A_1\cos\zeta + A_2 \cos^2\zeta + \cdots,
\end{eqnarray}
where the ellipses stand for higher order corrections and the coefficients are given by
\begin{eqnarray}
A_0 &=& \int d^2x ~ 2 \alpha_3 v^4 f_0^2 h_0^2,\\
A_1 &=& \int d^2x ~ v^2(h_0^2-f_0^2)\left(- m_2^2 +\alpha_4 v^2(h_0^2+f_0^2) \right),\\
A_2 &=& \int d^2x ~\alpha_3v^4(f_0^2-h_0^2)^2.
\end{eqnarray}
We omit $A_0$ since it is independent of $\zeta$. Note that $\cos\zeta$ and $\cos^2\zeta$ are odd and even under the $(\mathbb{Z}_2)_\mr{C}$
transformation ($\zeta \to -\zeta$), respectively.
The coefficient $A_1$ of $\cos\zeta$ only depends on $m_2$ and $\alpha_4$, whereas
the coefficient $A_2$ of $\cos^2\zeta$ only depends on $\alpha_3$. This is consistent with the fact that $\alpha_3$ does not 
break $(\mathbb{Z}_2)_\mr{C}$ because the $(\mathbb{Z}_2)_\mr{C}$ condition (\ref{eq:cond_2}) is independent of $\alpha_3$.
Namely, the $\cos^2\zeta$ term raises or lowers the tensions of both the $Z$ strings ($\zeta = 0,\pi$) keeping the degeneracy of them
while the $\cos\zeta$ term breaks the degeneracy.
The both terms do not change the tensions of the $W$-strings ($\zeta\sim \pi/2$).

For later use, we derive a sufficient condition that
the $W$-strings ($\zeta=\pi/2$) are heavier than the both $Z$ strings ($\zeta=0, \pi$)
up to $\mathcal{O}(m_2^4/v^4, \alpha_3^2,\alpha_4^2, \theta_\mr{W}^2)$.
It is equivalent to impose $\Delta T(\frac{\pi}{2}) > \Delta T(0,\pi)$:
\begin{align}
A_2 \pm A_1 < 0. 
\label{032346_1Feb20}
\end{align}
Since $A_2 = \alpha_3 \times (\text{positive constant})$, it implies that $\alpha_3$ should be negative with large $|\alpha_3|$.
For practical purposes, let us roughly estimate the condition Eq.~(\ref{032346_1Feb20}):
%\begin{align}
%  -\int d^2x ~ v^2(h^2-f^2)  \alpha_3 v^2(h^2-f^2) &>\int d^2x ~ v^2(h^2-f^2) \left| m_2^2 - ~\alpha_4 v^2(h^2+f^2) \right| \\
%&> \left|\int d^2x ~ v^2(h^2-f^2) \left(m_2^2 - ~\alpha_4 v^2(h^2+f^2) \right)\right| ,
%\end{align}
\begin{align}
\alpha_3 < - \frac{\displaystyle \left|\int dx^2\, v^2(h_0^2-f_0^2)\left(- m_2^2 +\alpha_4 v^2(h_0^2+f_0^2) \right)\right|}{
\displaystyle \int dx^2\, v^4(f_0^2-h_0^2)^2},
\end{align}
and by approximating the profile functions by $h_0(\rho)=1$ and $f_0(\rho) =\tanh(\rho v)$, we obtain an upper bound on $\alpha_3$:
\begin{align}
\alpha_3 \lesssim   -\left|3.69 \times \alpha_4 - 2.34 \times m_2^2/v^2 \right| + \mathcal{O}(m_2^4/v^4, \alpha_3^2,\alpha_4^2, \theta_\mr{W}^2).\label{215555_23Dec19}
\end{align}
Note that this bound is not rigorous
and valid only up to the leading order of $m_2,\alpha_3,\alpha_4$,
but provides a guide for parameter choices in numerical computations.

\subsection{Asymptotics of $Z$ strings}
We here investigate the asymptotic forms of the $Z$ strings at large distances.
Such an investigation is important to understand the dynamics and the stability of the string network in cosmology.
For general local vortices, e.g., the ANO vortex in the Abelian-Higgs model or superconductors,
an asymptotic form is given by an exponentially damping tale whose typical size is the mass scale of the model.
The stability of lattice structures of the ANO vortex (called as an Abrikosov lattice) is determined
by a ratio between sizes of tales of the scalar field and gauge field,
which is equal to the ratio of the scalar and gauge couplings.
On the other hand, for global vortices (e.g., axion strings),
the asymptotic form is given by a power-law tale because of the massless NG boson (axion particle).
This means that global vortices are much fatter than local ones
and that they have logarithmically divergent tensions.
In the present case for the 2HDM,
there are various mass scales in the mass spectrum including a massless CP-odd Higgs $A$ as shown in Sec.~\ref{172002_17Nov19},
so that the asymptotic form of the electroweak strings are quite non-trivial.
This situation is quite similar to a non-Abelian vortex in dense QCD \cite{Balachandran:2005ev,Nakano:2007dr,
Eto:2009kg,Eto:2009bh,Eto:2009tr}, 
see Ref.~\cite{Eto:2013hoa} as a review.
Here we follow the analysis of the asymptotic forms in Refs.~\cite{Eto:2009kg,Eto:2013hoa}.

Let us consider the $(1,0)$-string in the $(\mathbb{Z}_2)_\mr{C}$ symmetric case:
$m_2=\alpha_4=0$, hence $v_1=v_2\equiv v$.
By introducing new functions, the expression \eqref{eq:H(1,0)} can be rewritten as
\begin{align}
 H^{(1,0)}= \f{1}{2}v e^{i\varphi /2} ~e^{i\varphi \sigma_3/2 } \left(F(\rho) \bm{1} + G(\rho)\sigma_3\right),
\end{align}
where
\begin{equation}
 F(\rho)\equiv f(\rho)+ h(\rho), \h{2em} G(\rho) \equiv f(\rho)-h(\rho).
\end{equation}
Here, $F$ and $G$ are profile functions in the mass basis.
The former corresponds to the custodial singlet component $\chi^0$ in Eq.~\eqref{010632_18Mar20}
and the latter is the $\sigma_3$ component of the (split) custodial triplet, $\chi_3$.
We study the asymptotic forms of $F$, $G$ and $w^{(1,0)}$
at large distances compared to the inverses of the mass scales.
In this region, they are almost in the vacuum,
so that it is convenient to expand them around the vacuum as
\begin{equation}
 F(\rho)= F(\infty) + \delta F(\rho) = 2+ \delta F(\rho),
\end{equation}
\begin{equation}
 G(\rho) = G(\infty) + \delta G(\rho) = \delta G(\rho), 
\end{equation}
\begin{equation}
  w^{(1,0)}(\rho) = w^{(1,0)}(\infty) +\delta w(\rho) = \delta w(\rho).
\end{equation}
The linearized EOMs for $\delta F(\rho)$, $\delta G(\rho)$ and $\delta w(\rho)$ are given by
\begin{align}
  \left(\Delta_\rho - (m_{\chi^0})^2 - \f{1}{4 \rho^2}\right)\delta F(\rho) &=\f{1}{2\rho^2},\label{173942_8Feb20} \\
  \left(\Delta_\rho - (m_{\chi^3})^2 - \f{1}{4 \rho^2}\right)\delta G(\rho) &=\f{\delta w(\rho)}{\rho^2},\label{181148_8Feb20}\\
  \left(\partial_\rho ^2 - \f{1}{\rho}\partial_\rho -m_Z^2 \right) \delta w(\rho) &=m_Z^2 \delta G(\rho),\label{165310_8Feb20}
\end{align}
where $\Delta_\rho \equiv \f{1}{\rho}\partial_\rho (\rho \partial _\rho)$.
Eq.~(\ref{165310_8Feb20}) can be rewritten as
\begin{equation}
  \left(\Delta_\rho - m_Z^2 - \f{1}{\rho^2}\right)\delta \tilde{w}(\rho) =\f{m_Z}{\rho} \delta G(\rho)\label{181112_8Feb20}
\end{equation}
with $\delta \tilde{w}\equiv \delta w/(m_Z \rho)$.

Let us solve Eq.~(\ref{173942_8Feb20}).
The equation with the right hand side being zero has a solution proportional to $K_{1/2}(m_{\chi^0}\rho)$.
Here $K_{1/2}$ is one of the modified Bessel functions of the second class, which solves 
\begin{equation}
   \left(\Delta_\rho - m^2 - \f{n^2}{\rho^2}\right) K_n(m\rho) =0.
\end{equation}
Dealing with the right hand side iteratively, we obtain the asymptotic form of $F$ as
\begin{align}
 \delta F(\rho) \simeq q_F \sqrt{\f{\pi}{2 (m_{\chi^0})\rho}}e^{-(m_{\chi^0})\rho} -\f{1}{2 (m_{\chi^0})^2 \rho^2} + \mathcal{O}(\rho^{-4}) \simeq -\f{1}{2 (m_{\chi^0})^2 \rho^2} ,\label{180922_8Feb20}
\end{align}where $q_F$ is an integration constant
and we have used a fact that $K_n(r)$ with $0\leq r \leq 1$ behaves as $\sqrt{\f{\pi}{2r}} e^{-r}$ for $r \gg 1$.
The first term in Eq.~(\ref{180922_8Feb20}) is sufficiently small for $\rho \gg 1/m_{\chi^0}$.
Therefore, $\delta F(\rho)$ behaves as a power function $1/\rho^2$ for large $\rho$,
which leads to the logarithmic divergence in the string tension as discussed above.
This power-law tale is caused by the right hand side in Eq.~(\ref{173942_8Feb20}),
which is a source term generated by the massless particle $\pi^0$ (CP-odd Higgs $A$),
and a common feature for global vortices.

Let us next solve Eqs.~(\ref{181148_8Feb20}) and (\ref{181112_8Feb20}) using the iteration.
By setting the right hand sides in the equations to zero, we obtain
\begin{equation}
 \delta G(\rho) \simeq q_G  \sqrt{\f{\pi}{2 (m_{\chi^0})\rho}} e^{- \left(m_{\chi^3}\right) \rho},\label{183611_8Feb20}
\end{equation}
\begin{equation}
 \delta w \simeq q_Z \sqrt{\f{\pi m_Z\rho}{2}} e^{ -m_Z \rho} \label{121741_9Feb20},
\end{equation}
where $q_G$ and $q_Z$ are integration constants.
In realistic 2HDMs, the additional CP-even neutral Higgs $H$, as well as the SM Higgs $h$, is typically heavier than $m_Z$,
so that we take $m_{\chi^3} > m_Z$.
In this case, $\delta G(\rho)$ in Eq.~(\ref{183611_8Feb20}) is negligible
and ignoring the right hand side in Eq.~(\ref{181112_8Feb20}) is good approximation.
Thus, the leading expression Eq.~(\ref{121741_9Feb20}) is justified up to the sub-leading order of the iteration.
On the other hand, Eq.~(\ref{183611_8Feb20}) should receive the sub-leading iteration by substituting Eq.~(\ref{121741_9Feb20}) into Eq.~(\ref{183611_8Feb20}),
and we obtain
\begin{equation}
  \delta G(\rho) \simeq -q_Z \f{m_Z^2}{\left((m_\chi^3)^2-m_Z^2\right)} ~\sqrt{\f{\pi}{2 m_Z \rho}} e^{ -m_Z \rho}.\label{130029_9Feb20}
\end{equation}
Therefore, $ \delta G$ and $\delta w$ have the same exponential tales $e^{-m_Z \rho}$.
This result is a quite similar to a non-Abelian vortex 
in dense QCD \cite{Eto:2009kg,Eto:2013hoa},
instead of the ordinary ANO vortex in which each field has an exponential tale with each own mass scale.

The coefficient $q_Z$ is determined only by a numerical computation solving the EOMs.
Following the previous study of the $(1,0)$ string in Ref.~\cite{Eto:2018tnk},
we solve the EOMs for various range of $m_{\chi^3}/m_Z$ while fixing $m_{\chi^0}=m_h=125 ~(\text{GeV})$.
We fit the solutions by the asymptotic forms Eq.~(\ref{121741_9Feb20}) and (\ref{130029_9Feb20}) to obtain $q_Z$,
which are summarized in Tab.~\ref{131446_9Feb20}

We have investigated the asymptotic form of $(1,0)$-string for the $(\mathbb{Z}_2)_\mr{C}$ symmetric case.
The string has two tale structures;
one is the power-law tale of $F$ associating with the custodial singlet component, and 
the other is the exponential tale in $G$ and $w^{(1,0)}$ with the size of $1/m_Z$.
Therefore, the string has the logarithmic divergence in the tension while the $Z$ flux tube exponentially decays
as stated above.
The asymptotic forms of the $(0,1)$-string is the same as the above results thanks to the $(\mathbb{Z}_2)_\mr{C}$ symmetry.
In addition, those of the $W$-strings can be obtained by replacing $m_Z$ with $m_W$ in the above analysis.
On the other hand, when the potential does not have the $(\mathbb{Z}_2)_\mr{C}$ symmetry,
the situation could be more complicated.
However, the $Z$ strings still have the two structures that we explained above.
A quantitative discussion of the $Z$ strings without the $(\mathbb{Z}_2)_\mr{C}$ symmetry
requires a further analysis, which will be done elsewhere.

\begin{table}[tbp]
 \centering
 \scalebox{1.1}{
  \begin{tabular}{|c|cccc|}\hline
 $m_{\chi^3}/m_Z$ & 2 & 3 & 4 & 5 \\
 \hline 
 $q_Z(\delta G)$ & 2.60787 & 1.82783 & 1.65704 & 1.59833 \\
 $q_Z(\delta w)$ & 1.79159 & 1.61170 & 1.55712 & 1.53367\\
\hline
  \end{tabular}
  }
\caption{
 The obtained values of $q_Z$ by fitting the solution of the $(1,0)$ string.
 $q_Z(\delta G)$ and $q_Z(\delta w)$ denote the one using the asymptotic forms
 of $\delta G$ (Eq.~(\ref{130029_9Feb20})) and $\delta w$ (Eq.~(\ref{121741_9Feb20})), respectively.
 There is a good agreement between them for larger $m_{\chi^3}$ but not for smaller $m_{\chi^3}$.
 This is because the approximate expressions (\ref{121741_9Feb20}) and (\ref{130029_9Feb20}) are not valid
 for smaller $m_{\chi^3}$.
 }
\label{131446_9Feb20}
\end{table}

%%%%%%%%%%%%%%%%%%%%%%%%%%%%%%%%%%%%%%%%%%%%%%%%%%%%
%%%%%%%%%%%%%%%%%%%%%%%%%%%%%%%%%%%%%%%%%%%%%%%%%%%%

\section{Nambu monopoles}
\label{202730_25Feb20}
In this section, we study the Nambu monopole, which is a 't Hooft-Polyakov type magnetic 
monopole attached by the $Z$ strings in the 2HDM.
The static stable monopole was obtained under the restriction $(\mathbb{Z}_2)_\mr{C}$ symmetry in Ref.~\cite{Eto:2019hhf}.
In contrast, here, we will investigate the Nambu monopoles without the $(\mathbb{Z}_2)_\mr{C}$ symmetry.
\subsection{A point-monopole approximation}
Firstly, let us observe the monopole-string complex 
at large distance infinitely far from it.
Namely, we analytically deal with the $Z$ strings and Nambu monopole as infinitely thin and small objects.
An actual regular form will be constructed by a numerical relaxation method in the next subsection.

In the thin string limit, we replace, for instance in Eqs.~(\ref{eq:H(1,0)}) and (\ref{013304_3Dec19}), 
$f^{(1,0)}(\rho)$ by $f^{(1,0)}(\infty) = 1$, 
$h^{(1,0)}(\rho)$ by $h^{(1,0)}(\infty) = 1$, and $w^{(1,0)}(\rho)$ by $w^{(1,0)}(\infty) = 0$ 
for $\rho > 0$. Then, the singular $Z$-fluxes of the $(1,0)$- and $(0,1)$-strings are given by 
\begin{eqnarray}
F_{12}^{(1,0)} %= \partial_1 Z_2^{(1,0)} - \partial_2 Z_1^{(1,0)}
\to \cos^2 \beta\f{4\pi  \cos \theta_\mr{W}}{g}  \delta(x)\delta(y),\qquad
F_{12}^{(0,1)} %= \partial_1 Z_2^{(1,0)} - \partial_2 Z_1^{(1,0)}
\to - \sin^2 \beta\f{4\pi  \cos \theta_\mr{W}}{g}  \delta(x)\delta(y).
\end{eqnarray}
The Nambu monopole will play a role of a junction at which the two $Z$-fluxes are connected.
In Ref.~\cite{Eto:2019hhf}, we constructed the string-monopole complex at the $\tan\beta=1$ limit which
is quite special in a sense that the $(1,0)$- and $(0,1)$-strings can be transformed to each other 
by the $(\mathbb{Z}_2)_\mr{C}$ symmetry.
 With the aid of $(\mathbb{Z}_2)_\mr{C}$, we constructed the Nambu monopole
at $\tan\beta=1$ 
by acting the ``local'' $SU(2)_\mr{C}$ transformation 
that depends on the zenith angle $\theta$ in the real space. %on the $(0,1)$-string $H^{(0,1)}$.
We transformed the $(1,0)$-string which is put on the $z$-axis by
the local $SU(2)_{\rm C}$ transformation $U(\theta)=\exp\left(\frac{i\zeta(\theta)}{2}\sigma_1\right)$ 
with $\zeta(0) = 0$, $\zeta(\pi)=\pi$.
We have $U(\pi) = i\sigma_1$, which is nothing but the $(\mathbb{Z}_2)_{\rm C}$.
Thus, we obtained a configuration made of $(1,0)$- and $(0,1)$-strings on the positive and negative sides of the $z$-axes ($\theta=0,\pi$), respectively.

However, this construction cannot be used in the present case
because $(\mathbb{Z}_2)_\mr{C}$ is no longer a symmetry of Lagrangian 
and does not relate the two strings.
Instead, we provide a more general way to construct an asymptotic form of the Nambu monopole.
Let us start with $SU(2)_W$ adjoint composite fields 
$n^a_1$ and $n^a_2$ ($a=1,2,3$) 
 normalized to unity,\footnote{
This can be easily checked by using Fierz identities \cite{Nambu:1977ag}.
}
defined by 
\begin{equation}
n^a_i \equiv \frac{\Phi_i^\dagger \sigma^a \Phi_i }{\Phi_i^\dagger \Phi_i} \h{1em}(i=1,2).\label{224610_17Mar20}
\end{equation}
Note that they are analogues to a normalized adjoint scalar field 
for the 't Hooft-Polyakov monopole \cite{tHooft:1974kcl,Polyakov:1974ek}
because they determine the unbroken subgroup of the $SU(2)_W \times U(1)_Y$ gauge group.
In the following, we take $n_1^a = n_2^a$ 
because otherwise there is no unbroken subgroup and the photon becomes massive.
Since the $n_i$-fields are not well-defined at points in which $|\Phi_i|^2=0$,
we introduce the following adjoint field
which are well-defined even inside of the strings:
\begin{align}
 n^a \equiv& \f{ \sum_{i=1,2}\Phi_i^\dagger \Phi_i n_i^a}{\sum_{i=1,2}\Phi_i^\dagger \Phi_i}
  = \f{\Phi_1^\dagger \sigma^a \Phi_1 + \Phi_2^\dagger \sigma^a \Phi_2 }{\Phi_1^\dagger\Phi_1 + \Phi_2^\dagger \Phi_2}\label{234435_17Mar20} \\
=& \begin{cases}
    n_1^a & \text{for $\Phi_2=(0,0)^T$ and $\Phi_1\neq(0,0)^T$} \\
    n_2^a & \text{for $\Phi_1=(0,0)^T$ and $\Phi_2\neq(0,0)^T$} \\
    n_1^a (=n_2^a) & \text{otherwise}
\end{cases}
\end{align}
whose norm is unity.
Now, for a configuration with $n^a$ depending on $x^{1,2,3}$, 
the field strength for the electromagnetism and $Z$-boson are naturally defined by
\begin{eqnarray}
 F_{\mu\nu}^Z &\equiv&  -\cos \theta_\mr{W} n^a W^a_{\mu\nu} - \sin \theta_\mr{W} Y_{\mu\nu},\label{231032_19Nov19}\\
 F_{\mu\nu}^\mr{EM} &\equiv& - \sin \theta_\mr{W} n^a W^a_{\mu\nu} + \cos \theta_\mr{W} Y_{\mu\nu}\label{230623_19Nov19}, 
\end{eqnarray}
respectively.

Suppose there is a magnetic monopole at the origin surrounded by the vacuum of the 2HDM. Since the electromagnetic $U(1)_{\mr{EM}}$ is unbroken there,
the magnetic flux should spherically symmetrically spread from the monopole, as a usual Dirac or 't Hooft-Polyakov monopole.
Therefore, it is natural to impose a spherical symmetry on $n^a$,
%because the magnetic flux should be spherically symmetric at large distances
as described in Refs.~\cite{Nambu:1977ag,Eto:2019hhf}.\footnote{
Because $n^a$ is gauge dependent,
one can choose a gauge where $n^a$ is not spherical symmetric even when the magnetic flux is so.
}
Hence, we consider a configuration satisfying the so-called hedgehog structure:
\begin{equation}
 n^a = \f{x^a}{r} = (\sin \theta \cos \varphi , \sin \theta \sin \varphi, \cos \theta).
\end{equation}
The topological number of $n^a$ is unity similarly to the case of the 't Hooft-Polyakov monopole: 
$\f{1}{4\pi}\int_{r\to \infty} d\vec{S}\cdot \vec{n} = 1$.
The corresponding configuration of the original Higgs fields is given by
\begin{equation}
 \Phi_i^\mr{mon.} = v_i e^{i \phi_i} 
\begin{pmatrix}
 e^{-i \f{\varphi}{2}} \cos \f{\theta}{2}\\
  e^{i \f{\varphi}{2}} \sin \f{\theta}{2}
\end{pmatrix},\h{1em}(i=1,2)\label{005605_19Nov19},
\end{equation}
where we have used $ \Phi_i^\dagger \Phi_i = v_i^2$ except for the origin, and
$\phi_i$'s are arbitrary real functions but we have to choose $\phi_i$ in such a way that $\Phi_i$ becomes single valued.
Note that Eq.~(\ref{005605_19Nov19}) is quite similar to the configuration discovered by Nambu \cite{Nambu:1977ag} 
in the SM, except for the overall phase factor:
\begin{equation}
  \Phi_\mr{\mr{SM}} = v
\begin{pmatrix}
 \cos \f{\theta}{2}\\
  e^{i \varphi} \sin \f{\theta}{2}
\end{pmatrix}
\quad \xrightarrow[]{\theta \to \pi}\quad
v \begin{pmatrix}
 0\\
  e^{i \varphi}
\end{pmatrix},
\end{equation}
which describes a point-like magnetic monopole at the origin 
attached by an infinitely thin (non-topological) $Z$ string on $\theta=\pi$.
Since the 2HDM monopole is attached 
by {\it two topological $Z$ strings on the opposite sides} ($\theta = 0$ and $\pi$),
we take $\phi_i$'s as 
\begin{equation}
 \phi_1= - \f{\varphi}{2}, \h{1em} \phi_2=  \f{\varphi}{2},\label{005449_19Nov19}
\end{equation}
which ensures the single valued-ness of $\Phi_i$. 
This can be manifestly seen by the two-by-two matrix notation as 
\begin{equation}
  H^\mr{mon.} = \begin{pmatrix}
	       v_1 \sin \f{\theta}{2}& v_2 \cos \f{\theta}{2} \\
               -v_1 e^{i \varphi} \cos \f{\theta}{2} &   v_2 e^{i \varphi} \sin \f{\theta}{2}
	      \end{pmatrix} 
\quad \to \quad
\left\{
\begin{array}{ccl}
		  \begin{pmatrix}
	       0 & v_2 \\
               -v_1 e^{i \varphi} &   0
	      \end{pmatrix}& & \text{at}\ \theta = 0\\
		  \begin{pmatrix}
	       v_1 & 0 \\
               0 &   v_2 e^{i \varphi} 
	      \end{pmatrix}& & \text{at}\  \theta = \pi.
\end{array}
\right.
\label{163825_19Nov19}
\end{equation}
Eq.~(\ref{163825_19Nov19}) describes the $(1,0)$-string ($(0,1)$-string) on $\theta=0$ ($\theta=\pi$) 
up to the $SU(2)_W$ gauge transformation.\footnote{
This is clear when one acts the $SU(2)_W$ gauge transformation $H\to UH$ with $U$ satisfying
\begin{equation*}
 U|_{\theta=\pi}= 1_{2\times2} , \h{2em}U|_{\theta=0}= -i \sigma_2 .
\end{equation*}
}
This is a generalization of the one constructed in Ref.~\cite{Eto:2019hhf} for $\tan\beta=1$.

If we take $\phi_1 = \phi_2 = \varphi$ instead of Eq.~(\ref{005449_19Nov19}), we have another monopole-string configuration
quite similar to the Nambu monopole in the SM. There are two $Z$ strings on the negative side of the $z$ axis ($\theta = \pi$), but
they are $(0,1)$- and $(0,-1)$-strings. Thus, the configuration is not topologically protected at all, and we do not discuss it here.

As a next step, we consider the gauge fields $W_\mu$ and $Y_\mu$ in the presence of $H^\mr{mon.}$ [Eq.~(\ref{163825_19Nov19})].
They are determined to minimize the kinetic energy of the Higgs doublets,
\begin{equation}
\int d^3x~ \left( |D_i \Phi_1|^2+ |D_i \Phi_2|^2 \right) = \int d^3x~ \mr{tr}|D_i H|^2.\label{223311_17Mar20}
\end{equation}
After some lengthy calculations (see Appendix \ref{165251_19Nov19}), we obtain
\begin{equation}
g W_i^a + g' n^a Y_i = -n^a (\cos \theta + \cos 2\beta)\partial_i \varphi  - \epsilon^{abc}n^b \partial_i n^c.\label{223014_17Mar20}
\end{equation}
Note that the minimization condition for the Higgs fields can determine only the gauge fields corresponding to the broken generators.
Regarding $Y_i$ as an arbitrary function, we have
\begin{eqnarray}
F_{ij}^Z &=& %- g n^a W_{ij} ^a - g'Y_{ij} = 
\frac{\cos\theta_\mr{W}}{g}(\cos \theta +\cos 2\beta) \partial_{[i}\partial_{j ]} \varphi 
\label{eq:FZ}\\
F_{ij}^{\rm EM} &=& %-g' n^a W_{ij} ^a + gY_{ij} = 
\frac{\sin\theta_\mr{W}}{g}(\cos \theta +\cos 2\beta)  \partial_{[i}\partial_{j ]} \varphi
 + \frac{1}{\cos\theta_\mr{W}}Y_{ij},\label{eq:FEM}
\end{eqnarray}
where we have used identities
\begin{eqnarray}
 \vec n\cdot \left\{(\vec n\times \partial_i\vec n) \times (\vec n\times \partial_j\vec n)\right\} =  \frac{\epsilon^{aij}x^a}{r^3},\qquad
\vec n \cdot\left(\partial_{[i}\vec n \times \partial_{j]}\vec n\right) =  2\frac{\epsilon^{aij}x^a}{r^3},
\end{eqnarray}
\begin{equation}
 \sin \theta ~ \partial_{[i} \theta ~ \partial_{j]} \varphi = \f{\epsilon^{ija} x^a}{r^3}
\end{equation}
with $r^2=\rho^2+z^2$, $\cos \theta = z/\sqrt{\rho^2+z^2}$, and $\sin \theta = \rho/\sqrt{\rho^2+z^2}$.
Note $t_{[ij]} \equiv t_{ij} - t_{ji}$ for any tensor $t_{ij}$.
If the first term of $F_{ij}^{\rm EM}$ is present,
the configuration has a line singularity $\delta(x) \delta(y)$ on the $z$-axis,
which is inconsistent with the fact that $U(1)_{\mr{EM}}$ is unbroken.
Therefore, we must choose $Y_\mu$ to cancel such unphysical singular structures.  
An appropriate choice is given by
%Instead, $Y_\mu$ is determined so that the electromagnetic flux is not confined in the strings 
%(See Appendix \ref{165251_19Nov19}),
%resulting in
\begin{align}
 g W_i^a =& -\cos ^2 \theta_\mr{W} n^a (\cos \theta +\cos 2\beta)\partial_i \varphi,
 - \epsilon^{abc}n^b \partial_i n^c,\label{225144_19Nov19}  \\
g' Y_i =& -\sin ^2 \theta_\mr{W} (\cos \theta + \cos 2\beta)\partial_i \varphi.\label{225152_19Nov19}
\end{align}
Plugging these into Eqs.~(\ref{eq:FZ}) and (\ref{eq:FEM}), we get the final forms of the physical field strengths
\begin{align}
 F_{ij}^Z &=  \f{2\pi \cos \theta_\mr{W}}{g} \left( \f{z}{|z|} + \cos 2\beta \right) \epsilon_{3ij}\delta (x) \delta (y),\label{232727_19Nov19}\\
 F_{ij}^\mr{EM} &=  \f{\sin \theta_\mr{W}}{g} \epsilon^{aij} \f{x^a}{r^3},\label{231808_19Nov19}
\end{align}
where we have used the identity
\begin{eqnarray}
(\cos \theta + \cos 2\beta) \partial_{[i}\partial_{j ]} \varphi = 
2\pi \left( \f{z}{|z|} + \cos 2\beta \right) \epsilon_{3ij} \delta (x) \delta (y).
\end{eqnarray}

From Eq.~(\ref{231808_19Nov19}), it is clear that there is a magnetic flux from the origin in a spherical hedgehog form.
The total amount of the magnetic flux $\Phi_{\rm EM}$ can be calculated by integrating the flux density 
$B_i \equiv \f{1}{2}\epsilon_{ijk}F_{jk}^\mr{EM}$ as
\begin{equation}
 \Phi_{\rm EM} = \int d^3x~ \partial_i B_i =\f{4 \pi \sin \theta_\mr{W}}{g}.
\label{eq:EM_Sigma}
\end{equation}
Interestingly, the electromagnetic flux is independent of the ratio of the two Higgs VEVs, $v_1$ and $ v_2$.
This is understandable because the electromagnetic $U(1)$ remains unbroken, and the photon does not couple to the Higgs VEVs.

In addition, from Eq.~(\ref{232727_19Nov19}),
the $Z$-fluxes only exist on the $z$-axis as
\begin{eqnarray}
\Phi_Z\big|_{z > 0} &=& \int dx^2 \, F_{ij}^Z\big|_{z > 0} = \frac{4\pi\cos\theta_\mr{W}}{g} \cos^2 \beta = \Phi_Z^{(1,0)},\\
\Phi_Z\big|_{z < 0} &=& \int dx^2 \, F_{ij}^Z\big|_{z < 0} = - \frac{4\pi\cos\theta_\mr{W}}{g} \sin^2 \beta = \Phi_Z^{(0,1)},
\end{eqnarray}
flowing on the positive and negative sides of the $z$-axes, respectively, from the origin.
These amounts of the $Z$-fluxes agree with ones of the $Z$ strings given in Eq.~(\ref{eq:Z_fluxes}).
Therefore, the total amount of the $Z$-fluxes flowing from the monopole at the origin is independent of $\tan \beta$ as
\begin{eqnarray}
\Phi_Z= \int dx^3\, \partial_i B_i^Z = \Phi_Z\big|_{z > 0} - \Phi_Z\big|_{z < 0} = \frac{4\pi \cos \theta_\mr{W}}{g},
\end{eqnarray}
with $B_i^Z \equiv \frac{1}{2}\epsilon_{ijk}F_{jk}^Z$.
%Eqs.~(\ref{163825_19Nov19}), (\ref{225144_19Nov19}) and (\ref{225152_19Nov19}) actually 
%describe the infinitely small magnetic monopole attached by the two $Z$ strings.

It is worthwhile to demonstrate the topological current of $U(1)_a$ in the configuration.
The flux, corresponding to the winding of the $U(1)_a$ phase of the Higgs field,
is defined by 
\begin{equation}
\mathcal{A}_i \equiv \epsilon_{ijk} \partial ^j \mathcal{J}^k,\label{013151_11Mar20}
\end{equation}
\begin{equation}
\mathcal{J}_i \equiv -i ~\mr{tr} \left[H^\dagger D_i H  - (D_i H) ^\dagger H\right].
\end{equation}
Substituting Eq.~\eqref{163825_19Nov19} to this, we have
\begin{equation}
 \mathcal{A}_i = 8 \pi \sin ^2 \beta \cos ^2 \beta~ v_\mr{sum}^2 \delta_{i3} \delta(x)\delta(y).\label{183458_11Mar20}
\end{equation}
Importantly, $\mathcal{A}_i$ is topologically conserved, $\partial_i \mathcal{A}_i=0$, and independent of $z$.
This indicates that not only the string parts but also the monopole itself has the topological charge of $U(1)_a$.

\begin{figure}[ht]
\begin{center}
\includegraphics[width=10cm]{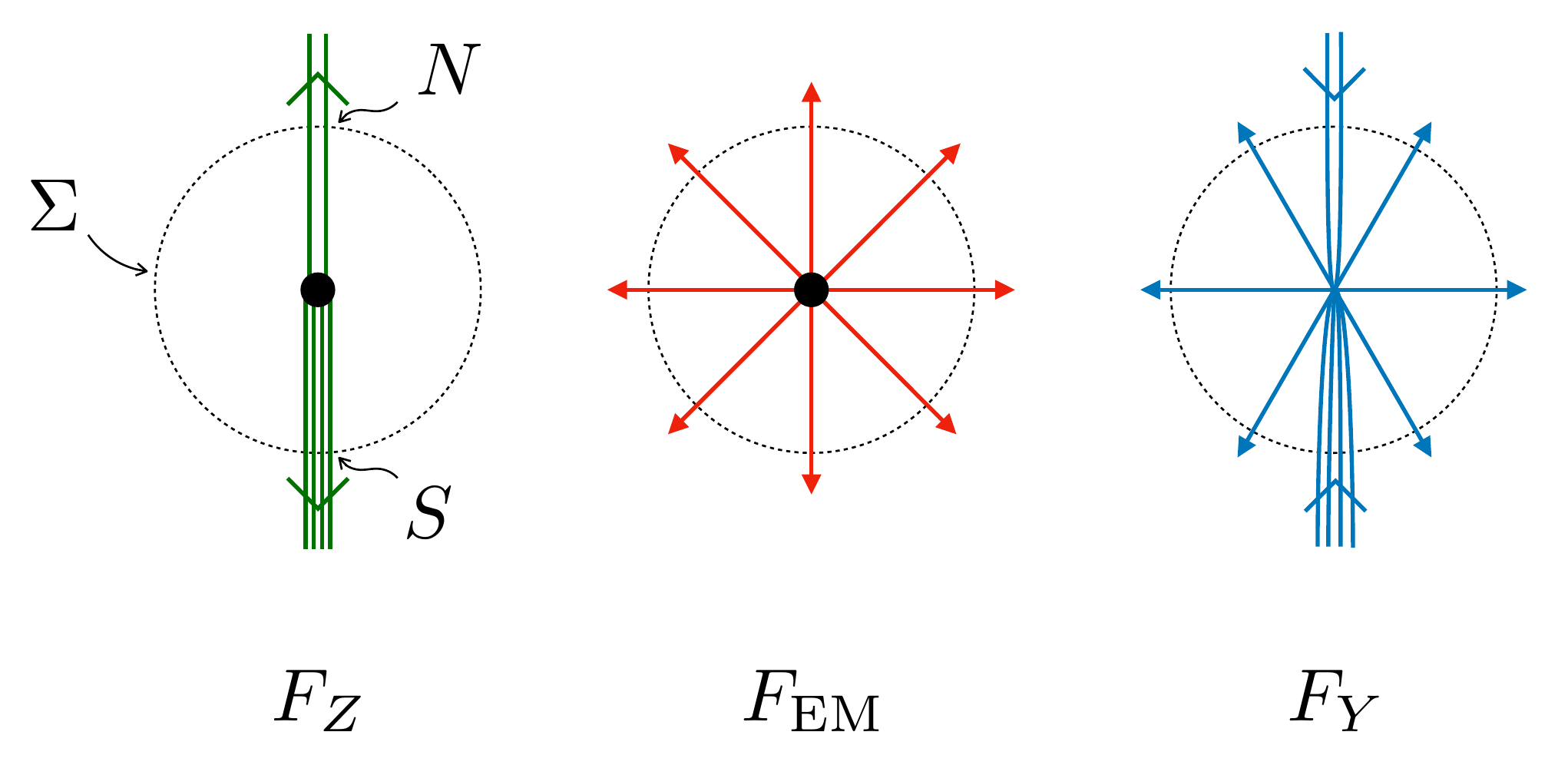}
\caption{The schematic pictures of the magnetic monopole in the 2HDM for $\tan\beta \neq 1$. The $(1,0)$-string passes through
the north pole while the $(0,1)$-string does through the south pole.}
\label{fig:schematic}
\end{center}
\end{figure}

Before closing this subsection, let us give a summary picture of the monopole in the 2HDM.
Let us denote the magnetic and $Z$ fluxes by $F_{\rm EM}$ and $F_Z$, 
respectively for simplicity. 
Similarly,
we denote the hypercharge- and $n$-magnetic fluxes by $F_Y$ and $F_n = \vec n \cdot \vec F_W$, respectively. 
Now consider a large sphere $\Sigma$ centered at the monopole, and let $S$ and $N$ be infinitesimally small regions at the
south and north poles, respectively. Eq.~\eqref{232727_19Nov19} tells
\begin{eqnarray}
F_Z\big|_{\Sigma-S-N} = 0,\quad
F_Z\big|_S =\f{4\pi \cos\theta_\mr{W}}{g}\sin^2\beta,\quad
F_Z\big|_N = \f{4\pi \cos\theta_\mr{W}}{g}\cos^2\beta.
\end{eqnarray}
Similarly, from Eq.~(\ref{231808_19Nov19}) we have
\begin{eqnarray}
F_{\rm EM}\big|_{\Sigma-S-N} = \f{4 \pi \sin \theta_\mr{W}}{g},\quad
F_{\rm EM}\big|_S = 0,\quad
F_{\rm EM}\big|_N = 0.
\end{eqnarray}
Combining these with $F_Y = -\sin\theta_\mr{W} F_Z + \cos\theta_\mr{W} F_{\rm EM}$ and $F_n = - \cos\theta_\mr{W} F_Z - \sin\theta_\mr{W} F_Y$, we observe
\begin{equation}
 F_Y\big|_{\Sigma-S-N} = \frac{4\pi\sin\theta_\mr{W}\cos\theta_\mr{W}}{g},
\end{equation}
\begin{equation}
F_Y\big|_{S} = -\frac{4\pi\sin\theta_\mr{W}\cos\theta_\mr{W}}{g}\sin^2\beta,\quad
F_Y\big|_{N} = -\frac{4\pi\sin\theta_\mr{W}\cos\theta_\mr{W}}{g}\cos^2\beta,
\end{equation}
and
\begin{equation}
 F_n\big|_{\Sigma-S-N} = - \frac{4\pi\sin^2\theta_\mr{W}}{g},
\end{equation}
\begin{equation}
F_n\big|_{S} = -\frac{4\pi\cos^2\theta_\mr{W}}{g}\sin^2\beta,\quad
F_n\big|_{N} = -\frac{4\pi\cos^2\theta_\mr{W}}{g}\cos^2\beta.
\end{equation}
Fig.~\ref{fig:schematic} shows the schematic pictures of the magnetic fluxes.
The magnetic monopole is the source for the $Z$ and magnetic fluxes. On the other hand, 
all the hypercharge-magnetic fluxes entering inside the strings go out through the sphere.
This is consistent with the fact that the hypercharge-magnetic field is divergenceless, so that
they cannot be terminated.
Note also that the net non-Abelian magnetic flux $F_n\big|_\Sigma = - \frac{4\pi}{g}$ equals to the one of the
conventional 't Hooft-Polyakov monopole.

\subsection{Ansatz for regular solutions}

The configuration constructed above is singular because it is just the asymptotic form at infinitely large distance.
In the next subsection, we will obtain regular solutions by numerically solving the equations of motion.
This subsection is devoted to prepartion for it. Namely, we provide an appropriate ansatz for the profile functions
of all the fields. The ansatz should be consistent with the asymptotic form obtained in the previous subsection, and
moreover a numerically low-cost ansatz is preferable.
Our starting point is rewriting the asymptotic gauge fields given in Eqs.~(\ref{225144_19Nov19}) and (\ref{225152_19Nov19})
in terms of the cylindrical coordinates as
\begin{align}
 W_\rho &=  \f{\cos \theta}{gr}\left(\sin \varphi ~ \f{\sigma_1}{2} - \cos \varphi ~\f{\sigma_2}{2}\right),\label{165003_2Dec19} \\
 W_\varphi &= \f{\rho}{gr}\left(\cos^2 \theta_\mr{W} \cos 2\beta  + \sin^2 \theta_\mr{W} \cos \theta \right)
 \left(\cos \varphi ~\f{\sigma_1}{2} + \sin \varphi ~\f{\sigma_2}{2}\right) \n \\
 & - \frac{1}{g} \left( \cos^2 \theta_\mr{W}  \cos \theta  \left(\cos\theta - \cos 2\beta \right) + \sin^2\theta \right) \f{\sigma_3}{2}, \\
W_z &=  \f{\sin \theta}{gr} \left(-\sin \varphi ~\f{\sigma_1}{2} + \cos \varphi ~\f{\sigma_2}{2}  \right),\\
Y_\varphi &= \frac{\sin ^2 \theta_\mr{W}}{g'} (\cos \theta - \cos 2\beta), \label{eq:Y_phi_inf}\\
Y_\rho &=0, \\
Y_z &=0.\label{165012_2Dec19}
\end{align}
This configuration has a cylindrical symmetry in the sense that 
it is invariant under the simultaneous rotations of the 2-dimensional real space $(x,y)$ and the internal space $(\sigma_1, \sigma_2)$.
In addition, it is also invariant under the simultaneous parity transformations: $\rho \to -\rho$ and $\sigma_{1(2)} \to -\sigma_{1(2)}$.

Let us introduce profile functions $\{u_1,u_2,u_3,u_4,b\}$ to smear the singularities
keeping the cylindrical symmetry as
 \begin{align}
 W_\rho (\rho,z) &= \f{u_1(\rho,z)}{g}\left(\sin \varphi ~ \f{\sigma_1}{2} - \cos\varphi ~ \f{\sigma_2}{2}\right),\label{083038_20Nov19} \\
 W_\varphi (\rho,z) &= \f{\rho~ u_2(\rho,z)}{g}\left(\cos \varphi~\f{ \sigma_1}{2} + \sin\varphi ~\f{\sigma_2}{2}\right)
                       +\f{\rho ~u_3(\rho,z)}{g} \f{\sigma_3}{2}, \label{083047_20Nov19} \\
  W_z (\rho,z) &= \f{ u_4(\rho,z)}{g}\left(- \sin \varphi ~ \f{\sigma_1}{2} + \cos\varphi~ \f{\sigma_2}{2}\right),\label{083053_20Nov19} \\
  Y_\varphi (\rho,z)&= \f{\sin^2\theta_\mr{W}}{g'} \rho ~b(\rho,z),\label{eq:Y_phi}\\
  Y_\rho &= 0,\\
  Y_z &=0 .\label{213624_2Dec19}
 \end{align}
Similarly, we make the ansatz for the Higgs fields by smearing the singular one given in Eq.~(\ref{163825_19Nov19}) as 
\begin{align}
 H &= \begin{pmatrix}
	       v_1 f_1(\rho,z) & -v_2 h_1(\rho,z) \\
               v_1 e^{i \varphi} f_2(\rho,z) &   v_2 e^{i \varphi} h_2(\rho,z)
	      \end{pmatrix}.\label{213601_2Dec19}
\end{align}
These profile functions should behave at $\rho \to \infty$ as
 \begin{align}
 u_1(\rho,z) &\to \f{\cos \theta}{r}, \\
 u_2(\rho,z) &\to \sin^2\theta_\mr{W}\f{\cos \theta}{r} + \f{\cos^2\theta_\mr{W}\cos 2\beta}{r} ,\\
 u_3(\rho,z) &\to - \cos^2\theta_\mr{W} \frac{z (\cos \theta -\cos 2\beta )}{\rho \sqrt{\rho^2+z^2}} -\frac{\sin \theta}{r},\\
 u_4(\rho,z) &\to  \f{\sin \theta}{r},\\
b(\rho,z) &\to \f{\cos \theta-\cos 2\beta}{\rho}, \label{122748_20Nov19}
 \end{align}
so that Eqs.~(\ref{083038_20Nov19})--(\ref{eq:Y_phi})
approach to the asymptotic forms in Eqs.~(\ref{165003_2Dec19})--(\ref{eq:Y_phi_inf}).
Similarly, we impose the following asymptotic behaviors at large distance on the rest of profile functions,
\begin{align}
  f_1(\rho,z) &\to \cos \f{\theta}{2},\\
  f_2(\rho,z) &\to \sin \f{\theta}{2},\label{122737_20Nov19}\\
  h_1(\rho,z) &\to \sin \f{\theta}{2},\\
  h_2(\rho,z) &\to \cos \f{\theta}{2},
\end{align}
so that Eq.~(\ref{213601_2Dec19}) approaches to Eq.~(\ref{163825_19Nov19}).
In addition, we impose boundary conditions on the profile functions at $\rho= 0$ as
\begin{equation}
  f_2=h_2=u_3=u_4=b=0,\label{021531_28Nov19}
\end{equation}
\begin{equation}
 \partial_\rho f_1=\partial_\rho h_1 =0 ,
\end{equation}
\begin{equation}
 \partial_\rho u_1 = \partial_\rho u_2 =0 \label{021512_28Nov19} ,
\end{equation}
where the Dirichlet conditions in the first line come from finiteness of the energy,\footnote{
To avoid singular energy density at $\rho=0$, we also need to keep the condition $u_1=u_2$ at $\rho=0$.
We ensure this additional condition by imposing it on the initial configuration at $\tau=0$. Then, the 
resulting configuration at any $\tau$ ($>0$) automatically satisfies $u_1=u_2$ at $\rho=0$.
}
and the Neumann conditions in the second line follow from smoothness of the Higgs field.
The last line is due to the parity symmetry under $\rho \to -\rho$, $\sigma_{1(2)} \to -\sigma_{1(2)}$. 
The energy density is indeed written down as follows,
\begin{eqnarray}
E = - K_W - K_Y - K_1 - K_2 + V,
\end{eqnarray}
with
\begin{align}
-2 g^2 K_W &=
(u_1'+\dot u_4)^2
+u_2'{}^2+\dot u_2^2
+u_3'{}^2+\dot u_3^2
-2 u_1 (\dot u_2 u_3 - u_2 \dot u_3)
+2 u_4 (u_2' u_3 -u_2 u_3' )\nonumber\\
&\ \ 
+(u_1^2 + u_4^2)(u_2^2+u_3^2)
+\frac{2 \left(-(u_1-u_2) \dot u_2 +\dot u_3 u_3+u_2' u_4+u_3 \left(u_1^2+u_4^2\right)\right)}{\rho}\nonumber\\
&\ \ + \frac{(u_1-u_2)^2+u_3^2+u_4^2}{\rho^2}
, \label{eq:K_W}\\
-2\frac{g_Z^4}{g'{}^2}K_Y & =
b'{}^2+\dot b^2+\frac{2 \dot b b}{\rho}+\frac{b^2}{\rho^2},\\
\frac{-4}{v_1^2}K_1 &= 
(2\dot f_1+u_1f_2)^2 
+ (2f_1'-u_4f_2)^2
+(2\dot f_2-u_1f_1)^2
+(2f_2'+u_4 f_1)^2 \nonumber\\
&\ \ 
+ \frac{(g'{}^2bf_1-g_Z^2(f_1u_3+f_2u_2))^2}{g_Z^4}
+ \frac{
\left\{\rho g_Z^2f_1u_2-f_2(g'{}^2\rho b + g_Z^2(\rho u_3+2))\right\}^2
}{\rho^2 g_Z^4},\\
\frac{-4}{v_2^2}K_2 &= 
(2\dot h_1-u_1h_2)^2 
+ (2h_1'+u_4h_2)^2
+(2\dot h_2+u_1h_1)^2
+(2h_2'-u_4 h_1)^2 \nonumber\\
&\ \ 
+ \frac{(g'{}^2bh_1-g_Z^2(h_1u_3-h_2u_2))^2}{g_Z^4}
+ \frac{
\left\{\rho g_Z^2h_1u_2-h_2(g'{}^2\rho b - g_Z^2(\rho u_3+2))\right\}^2
}{\rho^2 g_Z^4},\\
V &= 
\alpha_1 \left(
v_1^4 \left(f_1^2+f_2^2\right)^2
+2 v_1^2 v_2^2 (f_1 h_1-f_2 h_2)^2
+v_2^4 \left(h_1^2+h_2^2\right)^2
\right)\nonumber\\
&\ \ + \alpha_3 \left(
v_1^4 \left(f_1^2+f_2^2\right)^2
-2 v_1^2 v_2^2 (f_1 h_1-f_2 h_2)^2
+v_2^4 \left(h_1^2+h_2^2\right)^2
\right)\nonumber\\
&\ \ +
\alpha_2 \left(v_1^2 \left(f_1^2+f_2^2\right)+v_2^2 \left(h_1^2+h_2^2\right)\right)^2
+
\alpha_4 \left(v_1^4 \left(f_1^2+f_2^2\right)^2-v_2^4 \left(h_1^2+h_2^2\right)^2\right)\nonumber\\
&\ \ -
\mu_1^2 \left(v_1^2 \left(f_1^2+f_2^2\right)+v_2^2 \left(h_1^2+h_2^2\right)\right)
-
\mu_2^2 \left(v_1^2 \left(f_1^2+f_2^2\right)-v_2^2 \left(h_1^2+h_2^2\right)\right),
\end{align}
where we have used
$u' = \partial_z u$, $\dot u = \partial_\rho u$, and $g_Z^2 = g^2 + g'{}^2$.

We want to obtain the profile functions by solving the equations of motion
$\delta E/\delta X = 0$, 
where $X$ stands for the profile functions $X \in \{f_{1,2}, h_{1,2}, u_{1,2,3,4}, b\}$.
In general, however, even numerically it is not easy to solve such complicated differential equations.
So, instead of directly solving them,
here we make use of the relaxation method.
We introduce a fictitious time $\tau$ called as a flow time besides the real time $t$,
and evolve them by the following differential equations (flow equations):
 \begin{align}
  \partial_\tau X(\rho,z,\tau) = - \f{\delta E}{\delta X(\rho,z,\tau)}.\label{122816_20Nov19}
 \end{align}
 starting from some appropriate functions satisfying the boundary conditions given above
as an initial configuration at $\tau=0$. If the $\tau$-evolution converges, namely
$\partial_\tau X \to 0$ as $\tau$ evolves,
the convergent profile functions are nothing but the static solution of the original equations of motion.

Before going to solve the flow equations, however,
we should note that 
kinetic terms for $u_1$ and $u_4$ in Eq.~(\ref{eq:K_W}) are given by
\begin{equation}
\f{1}{2g^2}\left( {u_1'}^2 + \dot u_4^2 + 2 u_1' \dot u_4 \right),\label{180042_11Mar20}
\end{equation}
and that $\dot u_1^2$ and ${u_4'}^2$ are absent here.
%the quadratic derivative terms of $u_1$ and $u_4$ in Eq.~(\ref{eq:K_W}) are mixed
%whereas those of $u_2$ and $u_3$ are usual Laplacian.
Due to this, the flow equations for $u_1$ and $u_4$ are not genuine diffusion equations,
which are sometimes problematic because it is an obstacle to convergence.
%numerical instability is not suppressed in the equations.
To resolve this, we can use the gauge transformation,
$ W_i \to U \left( W_i + i g^{-1} \partial_i \right ) U^\dagger$ and
$ H \to U H$,  which
does not change forms of the ansatz
in Eqs.~(\ref{083038_20Nov19}), (\ref{083047_20Nov19}), (\ref{083053_20Nov19}),
and (\ref{213601_2Dec19}). Such $U$ is given by
\begin{equation}
 U = e^{i \omega\left(\sin\varphi ~\sigma_1 - \cos \varphi ~\sigma_2 \right)} 
 = \left(\begin{array}{cc}
 \cos \omega & -e^{-i\varphi}\sin\omega\\
 e^{i\varphi}\sin\omega & \cos \omega 
 \end{array}
 \right) \in SU(2)_W,
\end{equation}
which transforms the profile functions as
\begin{align}
 \begin{cases}
 u_1 \to u_1 + \dot\omega , \\
 u_2 \to u_2 + \f{2}{\rho} \omega +2 \omega u_3, \\
 u_3 \to u_3 -2 \omega u_2,  \\
 u_4 \to u_4 -\omega',\\
 f_1 \to f_1 \cos \omega - f_2 \sin\omega,\\
 f_2 \to f_1 \sin \omega + f_2 \cos\omega,\\
 h_1 \to h_1 \cos \omega  + h_2 \sin\omega,\\
 h_2 \to -h_1 \sin \omega  + h_2 \cos\omega,
 \end{cases}\label{103130_21Nov19}
\end{align}
where $\omega = \omega(\rho,z)$ is an arbitrary real function of $\rho$ and $z$.
By taking $\omega$ that satisfies
$\ddot \omega + \omega'' = \dot u_1 - u_4'$,
we can choose a gauge : $\dot u_1 - u_4'=0$.
Equivalently, we can simply add the following gauge fixing term
\begin{equation}
 \Delta E =\mr{tr} \left(\partial_\rho W_\rho+\partial_zW_z\right)^2
  =\f{1}{2g^2}\left(\dot u_1-u_4'\right)^2.
\end{equation}
Then, we find the normal quadratic terms for $u_1$ and $u_4$ in this gauge as
\begin{eqnarray}
E + \Delta E \supset \frac{1}{2g^2}(u_1'+\dot u_4)^2 + \frac{1}{2g^2}(\dot u_1 - u_4')^2 = \frac{1}{2g^2}\left(\dot u_1^2 + u_1'{}^2 + \dot u_4^2 + u_4'{}^2\right).
\end{eqnarray}
In summary, the improved flow equations to be solved are
\begin{align}
  \partial_\tau X(\rho,z,\tau) 
  = - \f{\delta (E+\Delta E)}{\delta X(\rho,z,\tau)}.
\end{align}

In the previously study \cite{Eto:2019hhf} by the present authors we experienced the similar numerical
computation but we did not make use of any symmetries to reduce numerical cost.
The relaxation scheme in Ref.~\cite{Eto:2019hhf} dealt with 20 fields (12 gauge fields $W_i^a, Y_i$ and 8 scalar fields in $H$)
which are dependent of the 3 spatial coordinates and the flow time $\tau$. Compared with Ref.~\cite{Eto:2019hhf},
the new relaxation scheme only includes 9 profile functions, and furthermore they are only dependent of 
the 2 spatial coordinates ($\rho$ and $z$) and  flow time $\tau$.

\subsection{Results of relaxation}

We show several results of solving the flow equations.
We will take the parameters in Lagrangian so that the $W$ strings are heavier than the $Z$ strings
because we want to study the Nambu monopole attached by the two $Z$ strings as is shown in Fig.~\ref{fig:schematic}.
Otherwise, the two $Z$ strings are unstable and would rapidly decay to the $W$ strings,
resulting in a single homogeneous (up to the $U(1)_{\mr{EM}}$ modulus) $W$ string without a monopole.
A rough condition for the $W$ strings to be heavier than the $Z$ strings was obtained in Sec.~\ref{172423_16Jan20}
[See Eq.~(\ref{215555_23Dec19})].

Throughout this subsection, we fix experimentally 
observed three parameters $\theta_\mr{W}, m_h, v_\mr{EW}$ as
\begin{equation}
 \sin^2\theta_\mr{W} = 0.23,~ m_h^2=(125 ~\mr{GeV})^2, ~ 2v_\mr{sum}^2 =v_\mr{EW}^2 = (246~\mr{GeV})^2.\label{012133_10Mar20}
\end{equation}
The other parameters such as the masses of the additional Higgs bosons and $\tan \beta$ are changed for several cases.

\subsubsection{$\tan \beta=1$ case}
Firstly, let us consider the case that the $(\mathbb{Z}_2)_\mr{C}$ is exact,
where the monopole is topologically stable because the tensions of the $(1,0)$- and $(0,1)$-strings are balanced.
This case was previously studied by the present authors in Ref.~\cite{Eto:2019hhf}
by constructing the stable solution based on the 3D simulation without any ansatz.
Therefore, we can check the consistency of our ansatz 
Eqs.~(\ref{083038_20Nov19})--(\ref{213624_2Dec19}) and (\ref{213601_2Dec19})
by comparing the result with that in Ref.~\cite{Eto:2019hhf}.

We take the following physical parameters in the 2HDM as \footnote{
As we stated above, $m_A$ vanishes because we impose the $U(1)_a$ symmetry.
The effect of $m_A \neq 0$ is discussed in Sec.~\ref{122632_28Nov19}.
}
\begin{equation}
  m_H^2= (400~ \mr{GeV})^2, ~m_{ H^{\pm}}^2= (400~ \mr{GeV})^2, ~ m_A^2=0,~ \tan \beta =1,
\end{equation}
and impose the alignment limit: $\cos(\beta-\alpha)=0$.
This is equivalent to choosing the parameters in the potential \eqref{173547_17Nov19} as
 \begin{align}
  m_1^2 = (0.719)^2 \times\f{v_\mr{sum}^2}{2}, ~ m_2^2 = 0, \\
\alpha_1 = 2.644,~ \alpha_2 = -1.193,~ \alpha_3 = 0, ~\alpha_4 = 0.
 \end{align}
In this choice, the Higgs potential has the custodial $SU(2)_\mr{C}$ symmetry,
and, therefore, the $(\mathbb{Z}_2)_{\rm C}$ condition (\ref{eq:cond_2}) is satisfied.
The energy of the $W$ strings are slightly heavier than those of the $Z$ strings because of $\theta_\mr{W}\neq 0$ as studied in Ref.~\cite{Eto:2018tnk}.

 %%%%%%%
\begin{figure}[htbp]
\begin{center}
\includegraphics[height=0.9\textheight]{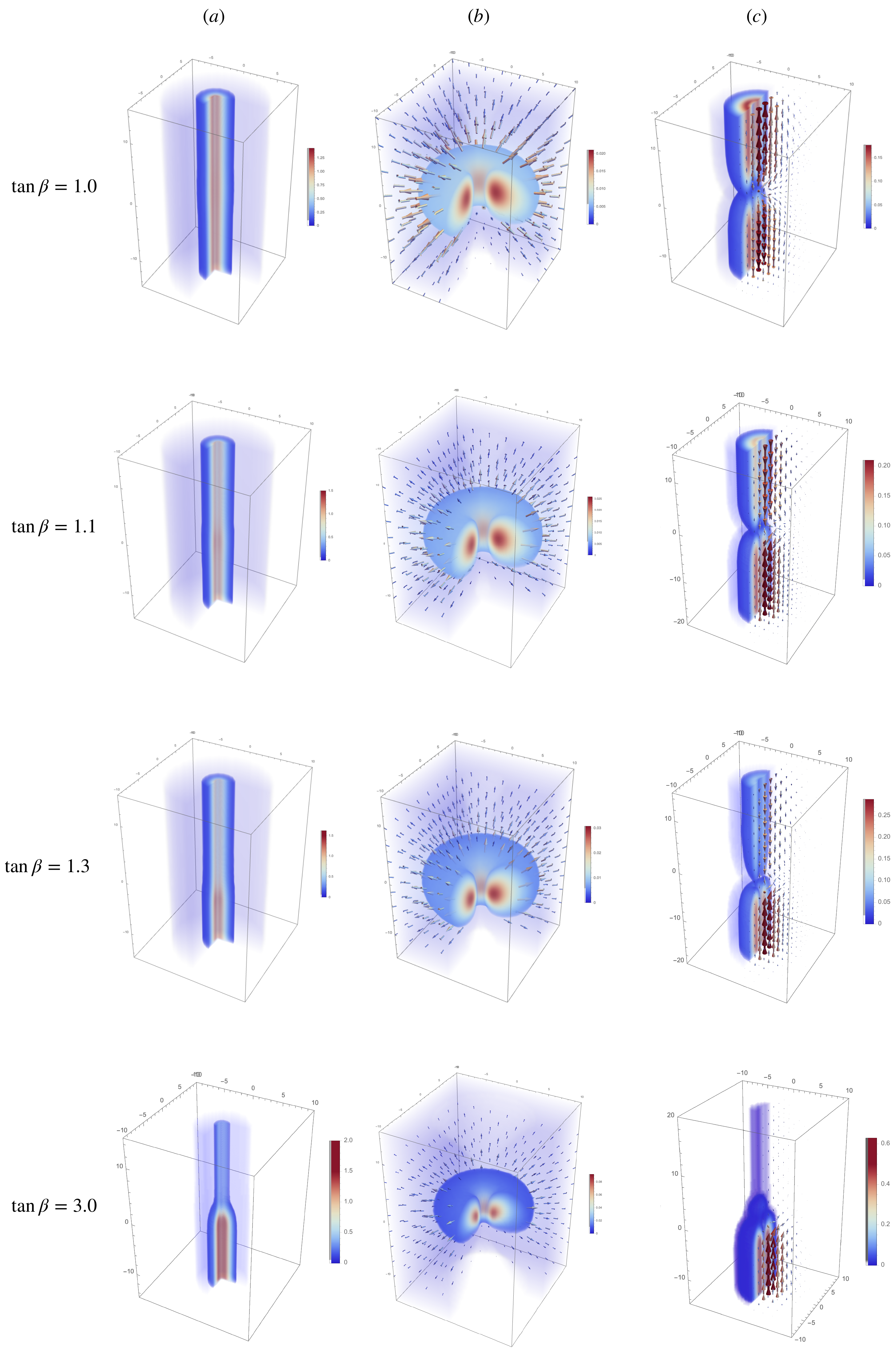}
\caption{
Plots for the Nambu monopole for the $(\mathbb{Z}_2)_\mr{C}$ symmetric and non-symmetric case.
%$\tan \beta=v_2/v_1$ indicates the deviation from the symmetric case
The symmetric case is $\tan \beta=1.0$.
In all plots, $\sqrt{(v_1^2+v_2^2)/2}=v_\mr{EW}/2$ is normalized to unity.
(a): Energy density. 
The color represents its value, where red is the largest and blue is the smallest. 
%There is a string-like object that contains the energy density along the $z$ axis.
(b): Magnetic flux. 
The direction of the arrows indicates that of the flux. 
Also, the color and size of the arrows indicate the flux density, where red is the strongest, blue is the weakest.
%We can see the existence of a magnetic monopole at the center.
(c): $Z$ flux. 
The direction, color and size of the arrows are the same as those for the magnetic flux. 
%The $Z$ flux flows upward and downward along the string from the monopole.
}
\label{032355_28Nov19}
\end{center}
\end{figure}
%%%%%%%
%

\begin{figure}[htbp]
\begin{center}
\includegraphics[width=0.9\textwidth]{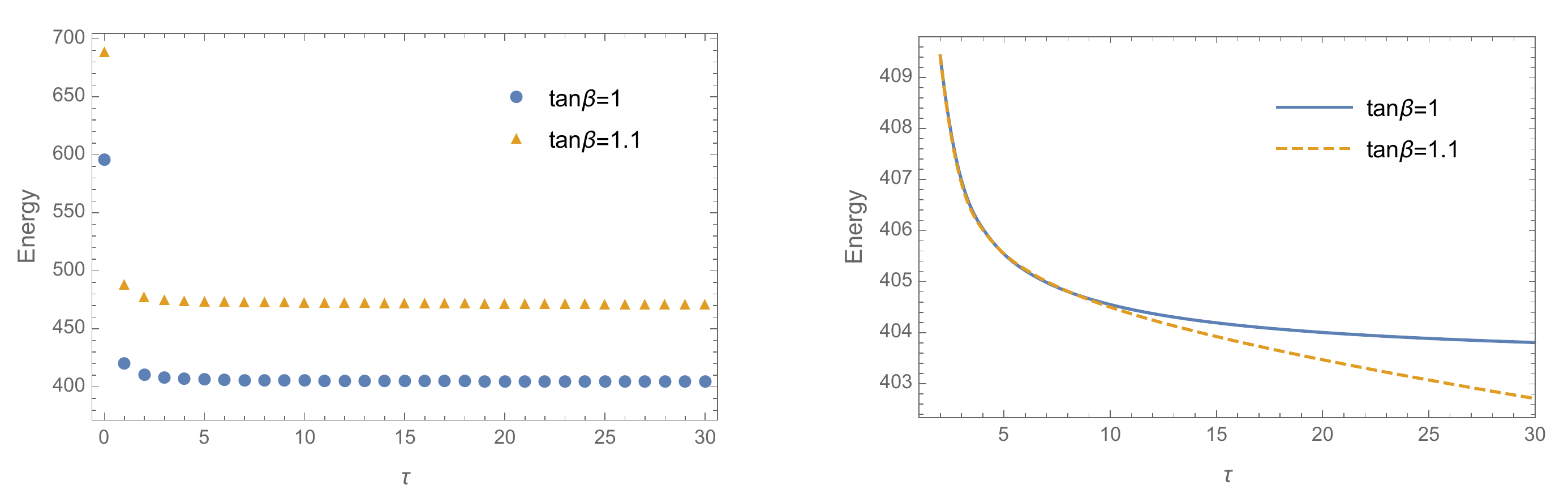}
\caption{
 Plot of the evoluion of the energy for the flow time $\tau$ 
for the $(\mathbb{Z}_2)_\mr{C}$ symmetric case ($\tan \beta=1.0$) 
and the non-symmetric case ($\tan \beta =1.1$).
 We evaluated the energy by integrating the energy density over 
$0<\rho<10$, $-15<z<20$ for $\tan \beta=1.1$ and $0<\rho<10$, $-15<z<15$ for $\tan \beta=1.0$, respectively.
 We adopt the unit $v_\mr{EW}/2=123~ \mr{GeV} \to 1$.
 The left panel shows the evolution for $0<\tau < 30$ 
($\tan \beta =1.0$ and $1.1$ is the blue circle and yellow triangle, respectively).
 The right one shows the interpolated energy plot for $2<\tau < 30$
($\tan \beta =1.0$ and $1.1$ is the blue solid line and yellow dashed line, respectively).
Here we have shifted the latter line to compare the shape with the former one. 
}
 \label{175139_24Jan20}
\end{center}
\end{figure}

The first line in Fig.~\ref{032355_28Nov19} shows the result at the flow time $\tau = 30$;
the energy density, magnetic flux and $Z$-flux from left to right.
The monopole attached by the two $Z$ strings on the opposite sides does not move and 
corresponds to a stable and static solution of the EOMs.
The blue dots in Fig.~\ref{175139_24Jan20} show the evolution of the energy for increasing $\tau$.
We can observe that it exponentially converges
and that the variation of the energy density per flow time is $\mathcal{O}(10^{-6})$ for $\tau \sim 30$,
so that we regard it as the end of the relaxation.

%%%%%%%
%%%%%%%

These results agree well with those of Ref.~\cite{Eto:2019hhf}.
Thus, the present ansatz Eqs.~(\ref{213601_2Dec19})-(\ref{213624_2Dec19}) are consistent 
and correct.
We emphasize that the ansatz is more economical and convenient in the sense of the numerical cost
compared to the full 3D computation.

\subsubsection{$\tan \beta \neq 1$ case}

Let us next consider the cases with $\tan \beta \neq 1$ where 
the $(\mathbb{Z}_2)_\mr{C}$ is explicitly broken.
In this case, the monopole cannot be static, since the monopole is attached
by the two $Z$ strings whose tensions are not equal each other.
As a result, there are no static monopole solutions of the genuine EOMs, and correspondingly
the energy does never converge in the relaxation.

As benchmark values, we take the following physical parameters:
%\begin{equation}
% \sin^2\theta_\mr{W} = 0.23,~ m_h^2=(125 ~\mr{GeV})^2, ~ 2v_{\rm sum}^2 =v_\mr{EW}^2 = (246~\mr{GeV})^2,\label{160448_28Feb20}
%\end{equation}
\begin{equation}
  m_H^2= (400~ \mr{GeV})^2, ~m_{ H^{\pm}}^2= (600~ \mr{GeV})^2, ~ m_A^2=0,~ \tan \beta =1.1,\label{160520_28Feb20}
\end{equation}
and impose the alignment limit: $\cos(\beta-\alpha)=0$.
Here $\theta_\mr{W}, m_h, v_\mr{EW}$ are the same as Eq.~\eqref{012133_10Mar20}.
These are equivalent to choosing the parameters in the potential as
\begin{align}
 m_1^2 = (0.719)^2 \times\f{v_{\rm sum}^2}{2}, ~ m_2^2 = 0, ~ \alpha_1 = 4.308, \\
  \alpha_2 = -1.193,~ \alpha_3 = -1.640, ~\alpha_4 = 0.254.
\end{align}
These choices satisfy the condition Eq.~(\ref{215555_23Dec19}).
We should note that the $(\mathbb{Z}_2)_\mr{C}$ symmetry 
is explicitly broken because $\alpha_4 \neq 0$ violates Eq.~(\ref{eq:cond_2}).

%%%%%%%
\begin{figure}[tbp]
\begin{center}
\includegraphics[width=0.8\textwidth]{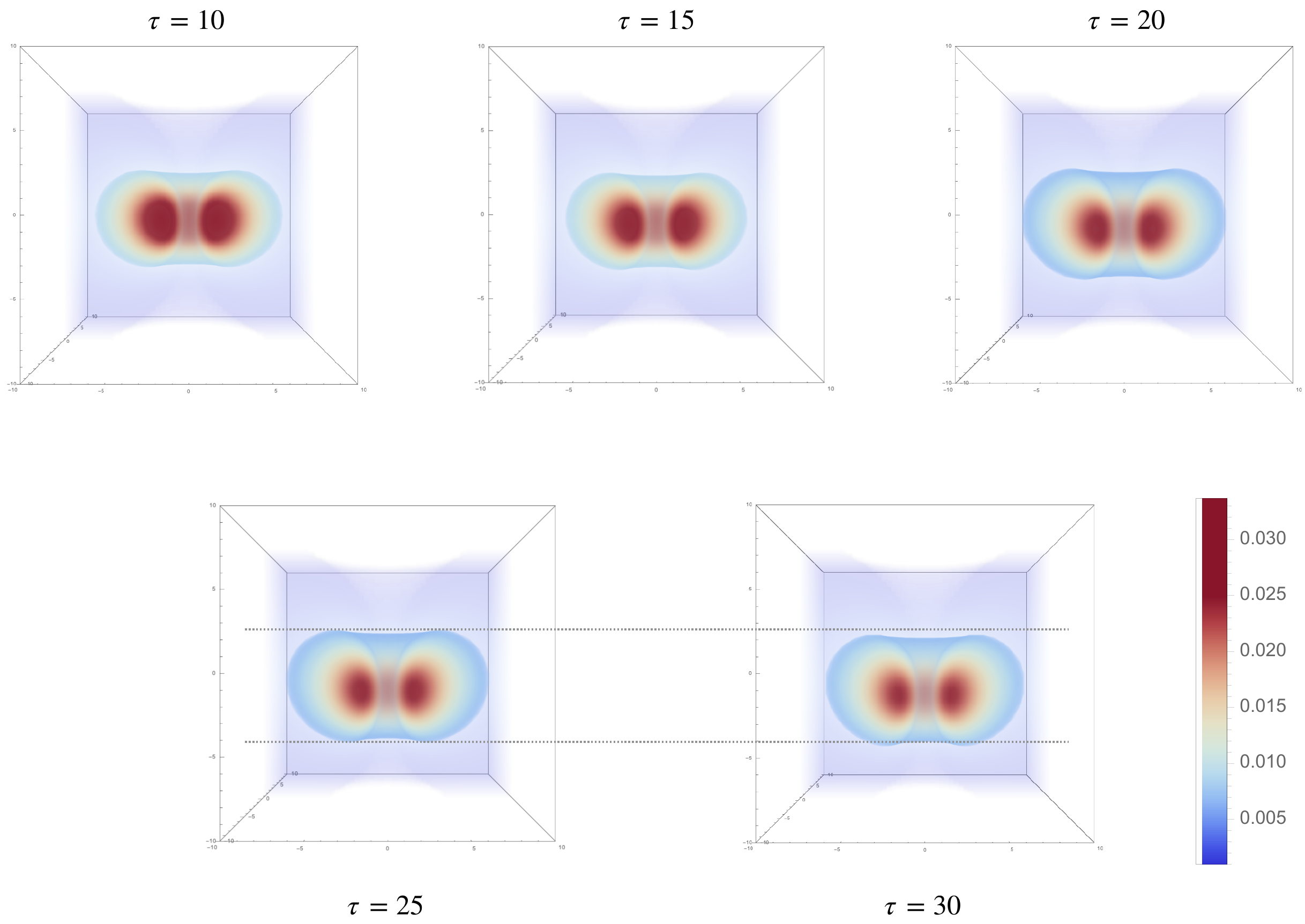}
 \caption{
 Snapshots of $|\vec{B}|^2$ at $\tau=10,15,20,25,30$.
 The plots are cut at $y=0$.
 Two horizontal dotted lines are shown to indicate the position of the monopole at $\tau=25$.
 It slowly moves down being pulled by the downside $Z$ string.
}
 \label{040945_28Nov19}
\end{center}
\end{figure}
%%%%%%%

We can observe that the monopole is pulled by the string as follows.
There are two stages of the $\tau$-evolution. The early stage is the period that
the energy exponentially decreases since the artificial initial configuration
rapidly releases the energy. After the first stage, the $\tau$-evolution goes into the
second stage in which the each piece of the configuration,
namely the monopole, or $(1,0)$- or $(0,1)$-strings, is not modified very much.
Instead, the monopole gradually shifts toward the heavier $Z$ string, which results in
the linear decreasing of the energy in the relaxation.
We can observe the two-stage $\tau$-evolution in several ways.
Fig.~\ref{040945_28Nov19} shows the evolution of the amplitude of the magnetic flux, $|\vec{B}|^2$.
For $0 < \tau \lesssim 25$, the size %structure 
and the density of the amplitude change significantly.
After $\tau \gtrsim 25$, they do not change, but slowly move down along the $z$-axis.
From this result, it is confirmed that
the monopole slowly moves and is pulled by the heavier $Z$ string unlike the previous case.
The orange dots in Fig.~\ref{175139_24Jan20} show the evolution of the energy for this case.
Unlike the previous case of $\tan\beta=1$, the evolution does not converge.
The first stage ($0 < \tau \lesssim 25$) is quite similar to the one of $\tan\beta=1$ but
the second stage ($\tau \gtrsim 25$) is peculiar to $\tan\beta\neq1$.
The energy continues to decrease by $\Delta z \times (T^{(1,0)}-T^{(0,1)})$
with $\Delta z$ being a distance the monopole moves.
As a result, the time dependence of the energy is linear in $\tau$ for the large flow time.
From the above two observations, we regard $\tau \sim 25$ as the end of the relaxation for the monopole.
%formed by $\tau=25$ and
%that it is only pulled by the $Z$ string for $\tau\geq 25$,
%which is consistent with the above observation based on the amplitude of the magnetic flux in Fig.~\ref{040945_28Nov19}.

The second line in Fig.~\ref{032355_28Nov19} shows the energy density, magnetic flux and $Z$-flux at $\tau=30$ from left to right.
Note that the amount of the $Z$ fluxes of the two strings are slightly different
since their ratio is given by $\tan^2\beta = 1.21$.
As a result, their energy densities are also different.
In addition, the monopole is not vertically symmetric, which is clear in Fig.~\ref{040945_28Nov19}.

%%%%%%%
\begin{figure}[tbp]
\begin{center}
\includegraphics[width=0.4\textwidth]{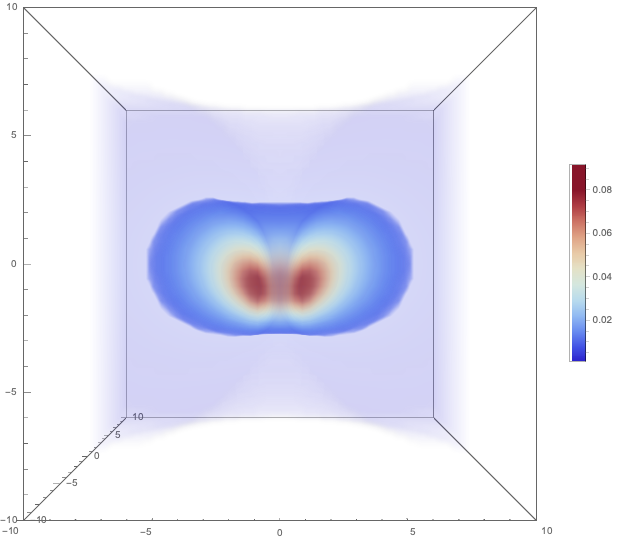}
 \caption{Cut of the amplitude of the magnetic flux at $y=0$ for $\tan \beta=3$.}
 \label{103538_26Jan20}
\end{center}
\end{figure}
%%%%%%%

We also compute for other values of $\tan \beta \neq 1$.
For examples, we consider two cases:
\begin{equation}
  m_H^2= (400~ \mr{GeV})^2, ~m_{ H^{\pm}}^2= (800~ \mr{GeV})^2, ~ m_A^2=0,~ \tan \beta =1.3,~ \cos(\beta-\alpha)=0
\end{equation}
and
\begin{equation}
  m_H^2= (400~ \mr{GeV})^2, ~m_{ H^{\pm}}^2= (1.8~ \mr{TeV})^2, ~ m_A^2=0,~ \tan \beta =3.0,~ \cos(\beta-\alpha)=0.\label{014126_10Mar20}
\end{equation}
The two-stage evolution in $\tau$ is qualitatively the same as the case of $\tan\beta=1.1$,
%The previous picture that the monopole is pulled by the $Z$ string does not change qualitatively,
so that we do not repeat the detailed explanations for the time evolution.
We only show the plots of $Z$ flux, the magnetic flux and the energy density 
in the third and fourth lines in Fig.\ref{032355_28Nov19}, respectively.
As expected, we can see that the amounts of the $Z$ fluxes of the two strings are significantly different.
Their ratio is $\tan^2 \beta = 1.69$ and $9$, respectively.
The energy densities are also different between the two strings,
however, the power-law tails coming from the winding of $U(1)_a$ phase are almost the same. 
%\red{(Can we show $U(1)_a$ flux at large distance?)}.
Furthermore, the shape of the magnetic flux is significantly distorted around the center, but it is not so at large distances.
This is clearly shown in Fig.~\ref{103538_26Jan20}.

\begin{figure}[tbp]
\begin{center}
\includegraphics[width=0.8\textwidth]{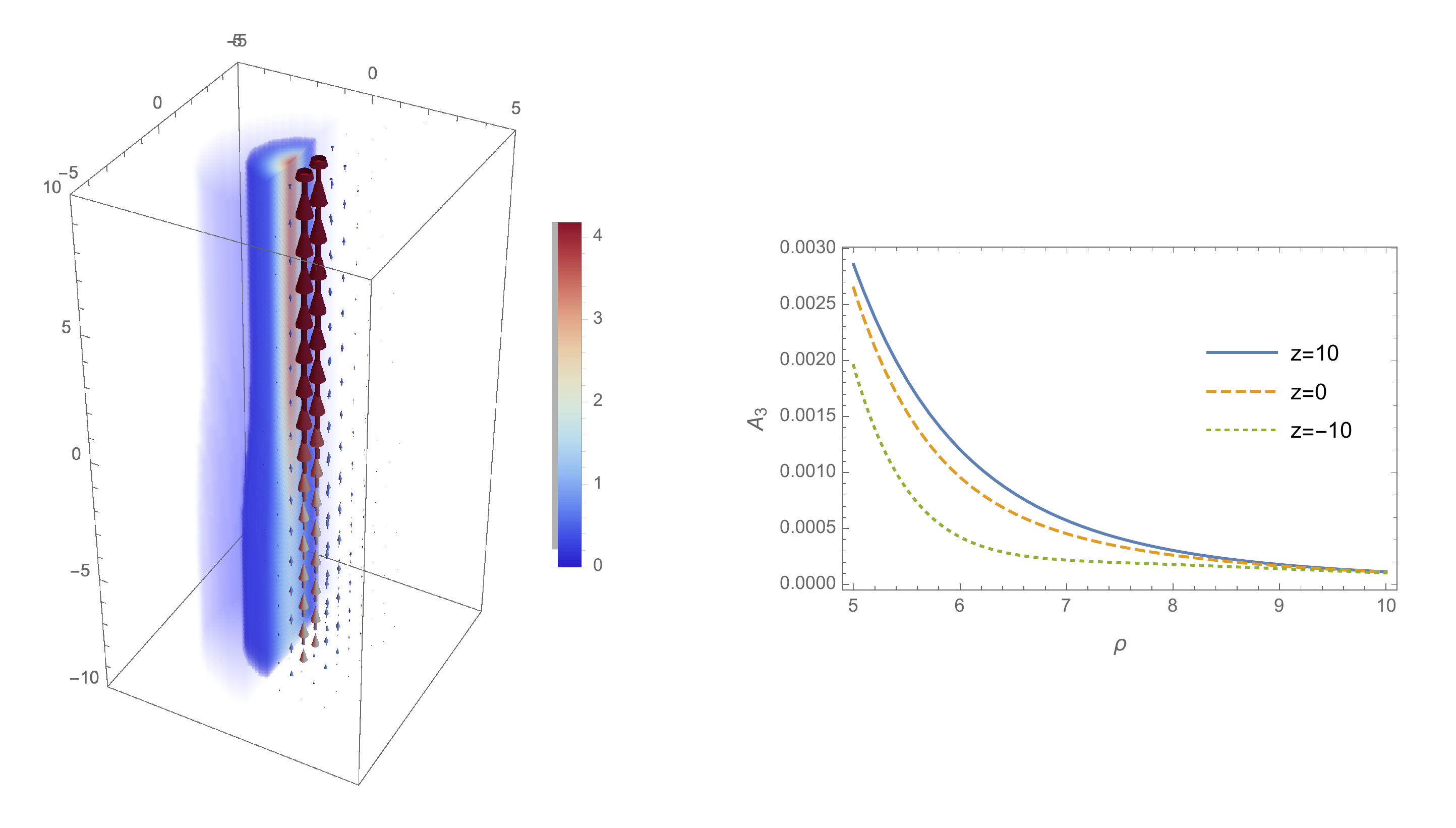}
\caption{
 Left panel: Plot of the $U(1)_a$ topological current $\mathcal{A}_i$.
 The direction, color and size of the arrows are the same as those for the magnetic flux.
 Right panel: Plot of $\mathcal{A}_3$ versus $\rho$ at $z=10,0,-10$.
 In both the panels, we take the parameter choice \eqref{014126_10Mar20} ($\tan \beta =3$ case).
}
\label{013553_10Mar20}
\end{center}
\end{figure}

Finally, we comment on the topological current of $U(1)_a$, $\mathcal{A}_i$, which corresponds to the winding of the $U(1)_a$ phase of the Higgs field.
The current is defined by Eq.~\eqref{013151_11Mar20}.
Importantly, $\mathcal{A}_i$ is topologically conserved, and independent of $z$ at large distances, see Eq.~\eqref{183458_11Mar20}.
Fig.~\ref{013553_10Mar20} shows the $U(1)_a$ current for the $\tan \beta=3$ case (Eq.~\eqref{014126_10Mar20}).
The density and width are different between the upper and lower sides in the vicinity of the string cores,
but the total flux integrated over a cross section $z= \text{const.}$ is always conserved.
This indicates that not only the string parts but also the monopole itself has the topological charge for $U(1)_a$.
%

%%%%%%%
%%%%%%%

%%%%%%%%%%%%%%%%%%%%%%%%%%%%%%%%%%%%%%%%%%%%%%%%%%%%
%%%%%%%%%%%%%%%%%%%%%%%%%%%%%%%%%%%%%%%%%%%%%%%%%%%%
\section{Monopole dynamics and radiation}
\label{105733_21Nov19}
We here discuss the real-time dynamics of the Nambu monopole without the $(\mathbb{Z}_2)_\mr{C}$ symmetry.
As shown in Sec.~\ref{202730_25Feb20}, the monopole moves along the strings being pulled by the heavier string.
It is difficult to analyze such a dynamics solving the EOMs of the gauge fields and the Higgs fields,
so that we deal with the monopole as a point-like object with a magnetic charge
and solve the classical mechanics.
As a result, the monopole is immediately accelerated to a velocity of order $1$ (the speed of light)
and emits electromagnetic radiation depending on the radius of curvature of the strings.
If the monopoles and strings existed in the early universe, 
the accelerated monopole collides to an anti-monopole with a kinetic energy of order $(\cos 2\beta)^{1/4} 10^8$ TeV,
whose remnants could be observed by the CMB anisotropy, primordial gravitational waves or the 21cm observation.

\subsection{Acceleration of Nambu monopole}
As we discussed in Sec.~\ref{172423_16Jan20}, the structure of the $Z$ strings consists of two parts;
an exponentially damping tail ($Z$ flux tube) corresponding to the $U(1)_Z$ winding 
and a fatter tail damping with a polynomial corresponding to the $U(1)_a$ winding, the latter of which leads to the log-divergent tension. 
%The polynomial part is irrelevant for the motion of the monopole along the string
%because 
The monopole attached with the two $Z$ strings is pulled by the difference of the tensions, which comes only from the exponential part. 
The log-divergent tension from the polynomial one is common on each cross section of the strings (See \eqref{202812_20Dec19})
because the $U(1)_a$ topological current is conseverd everywhere on the strings including the monopole.
Therefore, the polynomial parts are irrelevant as long as we consider the motion of the monopole along the strings.
We thus can approximate the exponentially damping structure (only $Z$ fluxes) of the strings to an infinitely thin one 
and the monopole to a point-like object 
keeping the width of the global $U(1)_a$ tails finite.

Based on this approximation, the dynamics of the point-like monopole is described by the following energy conservation law:
\begin{equation}
 \f{d}{dt} K + P_{\rm rad} = P_{\rm string}\label{203649_8Mar20}
\end{equation}
where $ K \equiv \gamma M$ is the kinetic energy of the monopole 
with static mass $M$ 
and $\gamma$ is the rapidity: $\gamma \equiv (1-u(t)^2)^{-1/2}$ with $u(t)$ being the velocity.
$P_{\rm rad}$ is the energy loss per unit time by the electromagnetic radiation from the monopole.
To calculate this, we just need to replace electric and magnetic variables in the well-known formula
of the radiation from an accelerated electric charge.
Then, we find
\begin{equation}
 P_{\rm rad} = \f{q_M^2}{4 \pi}
 \int  d\Omega ~
 \f{\left\{\vec{n} \times [( \vec{n}-\vec u)\times d\vec u/dt]\right\}^2}
  { (1- \vec{n}\cdot \vec u)^5} ,\label{134849_26Feb20}
\end{equation}
where $\Omega$ is the solid angle from the monopole point, $\vec{n}$ is a unit vector pointing from the position of the monopole to a point $\vec{x}$, and
$q_M$ is the magnetic charge of the monopole: $q_M= 4\pi \sin \theta_\mr{W}/g$.
%Here we have omitted terms in $\vec{B}$ that is subdominant ${\cal O}(|\vec x|^{-2})$.
On the other hand, $P_{\rm string}$ in Eq.~(\ref{203649_8Mar20})
is the energy gain per unit time from the string,
\begin{equation}
 P_{\rm string} = \Delta T u(t),\label{203637_8Mar20}
\end{equation}
where $\Delta T$ is the difference between the tensions of 
$(1,0)$ and $(0,1)$ strings.
Since the difference of the string tensions mainly comes from the $Z$-flux squeezed inside the
core of the strings, we estimate it as
\begin{eqnarray}
\Delta T \sim \frac{1}{2}\left[
\left(\frac{4\pi \cos\theta_\mr{W} \cos^2\beta}{g(\pi m_Z^{-2})}\right)^2 
-\left(\frac{4\pi \cos\theta_\mr{W} \sin^2\beta}{g(\pi m_Z^{-2})}\right)^2 
\right] (\pi m_Z^{-2})
= 2\pi v_{\rm EW}^2 \cos2\beta.
\end{eqnarray}

When the motion of the monopole is non-relativistic ($u\ll 1$),
the effect of the radiation $P_{\rm rad}$ is negligible and we have
\begin{equation}
 M \frac{du}{dt} = \Delta T u,
\end{equation}
which means that $u$ soon increases to of order $1$ with a time scale $ M/\Delta T$.
After that, we cannot ignore the relativistic breaking effect of the radiation.

Let us discuss the cases that the $Z$-stirings are straight and curved.
First, when the $Z$ strings are on a straight line, $\vec u$ and $d\vec u/dt$ are parallel.
Then, we have
\begin{eqnarray}
P_{\rm rad} = 
\frac{q_M^2}{4\pi} \left(\frac{du}{dt}\right)^2
\int d\Omega~ \frac{\sin^2\theta}{\left(1-u\cos\theta\right)^5}
=  \left(\frac{du}{dt}\right)^2 \frac{2q_M^2}{3}\gamma^6.\label{203626_8Mar20}
\end{eqnarray}
Combining Eqs.~\eqref{203626_8Mar20}, \eqref{203637_8Mar20} and \eqref{203649_8Mar20},
it is found that the monopole continues to accelerate emitting the radiation since $du / dt =0$ is not a solution.
The velocity approaches to the speed of light with infinite time.

\begin{figure}[tbp]
\begin{center}
\includegraphics[width=0.8\textwidth]{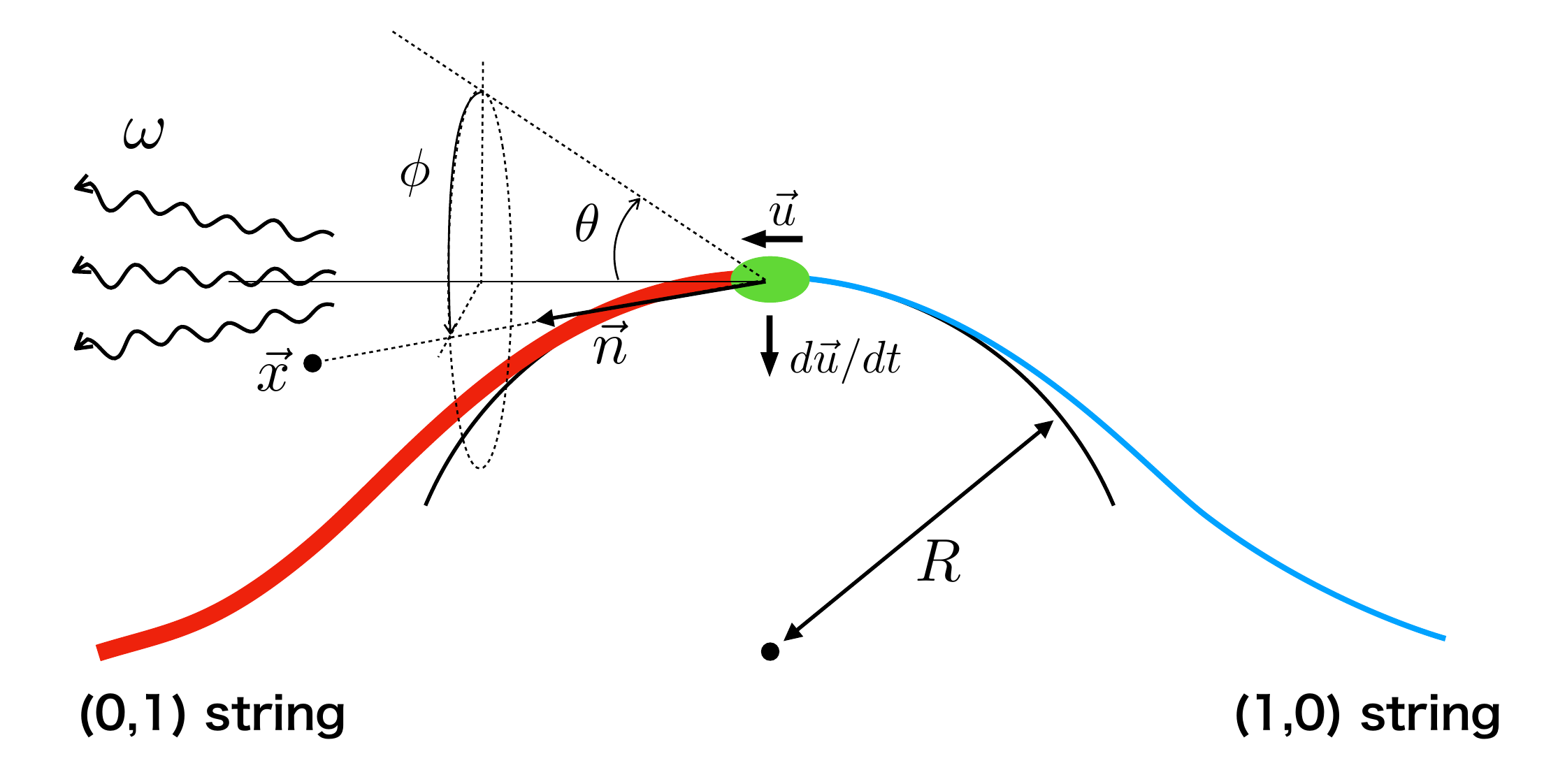}
 \caption{
 Accelerating Nambu monopole on the $Z$ strings.
 We approximate the string as a part of a circle with a radius $R$ of curvature. 
 The monopole generates the synchrotron radiation with a typical frequency $\omega$.
 }
\label{100728_21Feb20}
\end{center}
\end{figure}

The situation changes when the $Z$ strings bend.
Suppose the monopole runs along the string whose local curvature is $R$, see Fig.~\ref{100728_21Feb20}.
%%%%%%%
%%%%%%%
Even though the monopole speed reaches a maximum constant value, the monopole is still accelerated
toward the center of the circle, so the radiation does not stop.
It is a magnetic synchrotron accelerator.
The velocity is saturated by a maximum value $u_{\rm max} (\lesssim 1)$ 
when $P_{\rm rad} \sim P_{\rm string}$.
Since $\vec u \cdot d\vec u/dt = 0$, we have
\begin{eqnarray}
P_{\rm rad} &=& 
\frac{q_M^2}{4\pi} \left(\frac{du}{dt}\right)^2
\int d\Omega~ \frac{\left(1-u\cos\theta\right)^2 - \gamma^{-2}\sin^2\theta\cos^2\phi}{\left(1-u\cos\theta\right)^5} \nonumber\\
&=& \left(\frac{du}{dt}\right)^2
\frac{2 q_M^2}{3} \gamma^4 .
\end{eqnarray}
%Here we assume that the strings have a radius of curvature $R$
Then, we can roughly solve $P_{\rm rad} \sim P_{\rm string}$ as
\begin{equation}
q_M ^2 \f{u_{\rm max}^4\gamma^4}{R^2} \sim \Delta T u_{\rm max},
\h{2em} \therefore\ \  \left. \gamma\right|_{u_{\rm max}} \sim \left(\f{\Delta T R^2}{q_M^2} \right)^{1/4}, \label{092609_21Feb20}
\end{equation}
where we have used $du/dt = u^2/R$ and $u_{\rm max} \sim 1$.
When $u \sim 1$, the angular distribution has a strong peak at $\theta = 0$
similarly to the usual synchrotron radiation.
In addition, a typical frequency of a power spectrum of the synchrotron radiation is given by
\begin{equation}
 \omega \sim \gamma ^3 R^{-1} 
  \sim  \left(\f{\Delta T R^2}{q_M^2} \right)^{3/4}~ R^{-1}
  \sim  \left(\f{\Delta T }{q_M^2} \right)^{3/4} R^{1/2}.
%\sim  \f{\Delta T^{3/4} R^{1/2}}{q_M^{3/2}}
\label{175340_21Feb20}
\end{equation}
Note that $\omega$ becomes larger for the larger radius  $R$ of the curvature
because the maximum velocity $u_{\rm max}$ becomes larger, 
producing the larger blue-shift effect.

\subsection{Cosmological monopole collider}
Let us consider the acceleration of the monopole presented above in the early universe.
To discuss the cosmological scenario of the strings and the monopole, we here give the following two assumptions.
The first assumption is that the difference of the tensions between (0,1) and (1,0) strings is not very large compared to the electroweak scale $v_\mr{EW}^2$.
The second is that the difference of the string tensions between the $W$ and $Z$ strings is less than $v_\mr{EW}^2$.
From these assumptions, we can estimate the reconnection probabilities of our strings.
A pair of the strings of the same kind can reconnect with a probability of order unity as usual for global strings.
The reconnection of a pair of (1,0) and (0,1) strings creates a pair of a monopole and an antimonopole on the reconnected strings,
and thus its probability depends on the ratio of the difference of the tensions of $W$ and $Z$ strings to the string kinetic energy 
and is of order unity from the second assumption.%
\footnote{
These estimates are based on Refs.~\cite{Hashimoto:2005hi,Eto:2006db} in which a model closely related to our case with the $U(1)_Y$ coupling $g'$ switched-off is studied, 
and so are presumably valid in our case as well.
}
Therefore, we qualitatively deduce \cite{Kibble:1976sj,Kibble:1980mv,Vilenkin:1981kz,Kibble:1984hp,Bennett:1985qt,Bennett:1986zn}
%that if we ignore the magnetic fluxes of the monopoles 
%(or switch off the $U(1)_Y$ coupling $g'$), 
that the strings produced during the electroweak phase transition by the Kibble-Zurek mechanism \cite{Kibble:1976sj,Zurek:1985qw} would form a complex network consisting of (1,0) and (0,1) strings and the monopoles, 
in which the typical scale is the horizon scale $d_H$,
{\it i.e.}, there are a few strings and monopoles per the Hubble horizon scale $d_H$.%
\footnote{
These issues should be confirmed by numerical simulations, which are very challenging studies beyond the scope of this paper.
}

After the temparature of the universe decreases to the difference of the $Z$ string tensions,
the difference becomes relevant 
and hence the monopoles on the string network start to move along the strings as analyzed in the last subsection.
Thus, the radius $R$ of curvature of the strings in Eq.~\eqref{092609_21Feb20} is naturally taken as the Hubble radius:
\begin{equation}
 R \sim d_H \sim \f{M_P}{\sqrt{g_\ast}\, T_{\rm th}^2}.
\end{equation}
Here, $g_\ast$ is the number of the effective degrees of freedom in the thermal bath
and $T_{\rm th}$ is their temperature, which is assumed to be almost constant.
Substituting $q_M \sim 1$, $g_\ast \sim 10^2$, $ T_{\rm th} \sim v_{\rm EW}$ and
$\Delta T \sim  v_{\rm EW}^2\cos 2\beta$
into Eq.~\eqref{092609_21Feb20},
we obtain the maximally accelerated rapidity of the monopole as
\begin{align}
 \gamma \sim \left(\f{\Delta T R^2}{q_M^2} \right)^{1/4}
% & \sim \left( v_\mr{EW}^2 \f{M_P^2}{g_\ast v_\mr{EW}^4} \right)^{1/4} \\
  \sim \left( \f{\cos 2\beta ~M_P^2}{q_M^2 g_\ast v_{\rm EW}^2} \right)^{1/4} 
% & \sim 10^{(38-2-4)/4}\\
 \sim (\cos 2\beta)^{\f{1}{4}}~ 10^8.
\end{align}
Furthermore, the typical frequency of the synchrotron radiation is
\begin{equation}
 \omega \sim  \gamma ^3 R^{-1}
%  \sim 10^{24} \f{\sqrt{g_\ast} T_{th}^2}{M_P}.
%  \sim 10^{24} \f{10 * 10^4 }{10^{19}}~ \mr{GeV}.
  \sim (\cos 2\beta)^{\f{3}{4}}~10^{10}~ \mr{GeV}.
\end{equation}

After the acceleration, the monopole collides to an anti-monopole on the $Z$ strings.
The kinetic energy of the monopole is given by 
\begin{equation}
 K_{\rm max} = \gamma M \sim (\cos 2\beta)^{\f{1}{4}}~ 10^{11} ~\mathrm{GeV},
\end{equation}
where we have used $M \sim 1 ~ \mathrm{TeV}$.
After the collision, they would annihilate and produce radiations or particles
with the energy of the order $10^{11} ~\mathrm{GeV}$ for $(\cos 2\beta)^{1/4} \sim 1$.
Interestingly, this is a very high-energy phenomenon,
which we call as {\it the cosmological monopole collider} (CMC).
Remnants of the collisions and the synchrotron radiation could be observed
by the CMB anisotropy, primordial gravitational waves or the 21cm observations.

Before closing this section, we emphasize the difference 
between CMC and other monopole and string systems.
Let us consider three examples;
the Nambu monopole in the SM \cite{Nambu:1977ag}, 
a confined GUT monopole in the Langacker-Pi mechanism \cite{Langacker:1980kd}
and cosmic necklaces \cite{Berezinsky:1997td,Siemens:2000ty} consisting of string networks with confined monopoles.
The first one is a magnetic monopole attached by a non-topological $Z$ string from one side.
The second one appears when $U(1)_\mr{EM}$ is spontaneously broken in the early universe
and is also pulled by a $U(1)_\mr{EM}$ string ($U(1)_\mr{EM}$ flux tube) from one side.\footnote{
For example, consider the following symmetry breaking: 
$SU(5)\to SU(3)_\mr{Color} \times U(1)_\mr{EM} \to SU(3)_\mr{Color}$.
The $U(1)_\mr{EM}$ string has two end points with the GUT monopole and anti-monopole
because $\pi_1(SU(5)/SU(3)_\mr{Color})=0$.
}
Both of them could accelerate being pulled by the strings.
However, the strings are not topological and do not form any string networks.
As a result, there were only small pieces of string segments
with end points of the monopole and the anti-monopole in the universe.
Therefore, the monopoles in the first and second examples collide to the anti-monopoles before sufficient acceleration 
and disappear without any relevant radiation.
On the other hand, the third one is similar to our case in the sense that the strings form a complex network with the monopoles.
But it differs in that there is no electromagnetic fluxes spreading from the monopoles 
and that the monopole is not pulled by the strings, which results in no acceleration.
Thus, CMC is peculiar to the Nambu monopole in the 2HDM,
in which the monopole is pulled by the {\it topological} $Z$ string.

\section{Discussion and conclusion}
\label{122632_28Nov19}
In this paper, we have investigated the dynamics of the Nambu monopole in the 2HDM.
Firstly we have studied the electroweak strings 
and found that asymptotic behaviours of profile functions 
of gauge and scalar fields
decay with the mass of Higgs fields,
unlike vortices in superconductors.
We have derived the condition \eqref{215555_23Dec19} that the $Z$ strings are stabler than the $W$ strings, equivalently 
that the Nambu monopoles are stable on the $Z$ string.
Next we have given an asymptotic form of the monopole in the case without the $(\mathbb{Z}_2)_\mr{C}$ symmetry
based on the point-like approximation for the monopole.
The two $Z$ strings attached to the monopole have different $Z$ fluxes with its ratio being $\tan^2 \beta$
because of the absence of $(\mathbb{Z}_2)_\mr{C}$.
On the other hand, the magnetic flux spreading from the monopole does not depend on $\tan \beta$
and is spherically symmetric at large distances.
Based on the asymptotic form, we have presented a cylindrical ansatz describing the regular monopole configuration.
It is much more convenient than the full 3D one that we used in the previous work \cite{Eto:2019hhf}.
After that, we have shown results of the relaxation method using the ansatz for several parameter choices.
In the $(\mathbb{Z}_2)_\mr{C}$ symmetric case, $\tan \beta=1$, the monopole does not move and is static solution of the EOM,
which is consistent with the previous work \cite{Eto:2019hhf}.
On the other hand, the monopole moves on the string being pulled by the heavier one 
for the non $(\mathbb{Z}_2)_\mr{C}$ symmetric case, $\tan \beta \neq 1$.
See Figs.~\ref{032355_28Nov19} and \ref{040945_28Nov19}.
In the last section, we have analyzed the real time dynamics of the monopole based on the point-like approximation.
The monopole accelerates by the string emitting electromagnetic radiations
like a synchrotron accelerator.
We have considered the CMC, a high-energy phenomenon in the early universe that
an accelerated monopole collides to an anti-monopole with kinetic energy $\sim (\cos 2\beta)^{1/4} 10^{11}$ GeV.

Let us comment on the CMC and its phenomenological implication.
Interestingly, the center-of-mass energy of the collision event at the CMC 
is around $10^{11}$ GeV for $ (\cos 2\beta)^{1/4} \sim 1$,
which is much higher than those that our collider experiments can reach today.
By the collision, heavy particles with masses of the order $10^{11}$ GeV can be produced as in an ordinary synchrotron collider.
Such remnants could remain as fluctuations of matter distributions in the present universe 
and be observed by the CMB anisotropy, primordial gravitational waves and the 21cm observations.
Therefore, the CMC can be a tool built in nature to probe high energy physics beyond the SM
such as inflation models and GUTs.
This situation is quite similar to the so-called cosmological collider \cite{Arkani-Hamed:2015bza,Chen:2009zp,Chen:2009we,Noumi:2012vr,Baumann:2011nk}.%
\footnote{Needless to say, our terminology ``cosmological monopole collider'' is an imitation of the cosmological collider.}

In this paper, the $U(1)_a$ symmetry is imposed in the Higgs potential, 
so that the stability of the $Z$ strings is topologically protected.
As we stated in Sec.~\ref{172002_17Nov19}, however, it should be explicitly broken 
by switching on $m_3, \alpha_5$ to make the CP-odd Higgs boson massive.
The effect of $m_3, \alpha_5 \neq 0$ is discussed by some of the present authors in Refs.~\cite{Eto:2018hhg,Eto:2018tnk}.
Similarly to the axion string and domain wall associated with the $U(1)_\mr{PQ}$ symmetry, 
which is explicitly broken by the axial anomaly,
the $Z$ strings and the monopole are attached by one or two domain walls depending on the values of $m_3, \alpha_5$.
Therefore, our observation that the monopole moves along the string network would be more complicated,
that is, the monopole is pulled by the string, 
and both of them are pulled by the wall.
While our study is justified when the tension of the wall is small compared to $\Delta T$,
in general, we have to consider two directions of the acceleration of the monopole,
and the CMC is no longer a simple synchrotron accelerator.
A further quantitative study is needed to estimate the energy of the CMC, 
which is left for future work.

We discuss the abundance of the monopoles in the present universe,
which is important to see whether the monopoles dominate the energy density of the universe (cosmological monopole problem) or not.
As we stated above, the monopole accelerates by the CMC
with a typical time scale $\sim M/\Delta T$.
After the acceleration, the monopole and anti-monopole collide and annihilate immediately,
and hence the monopoles would not remain abundant
unless $\Delta T$ is unnaturally small,
i,e, fine tuned to be a small value.\footnote{
Even if such a fine tuning is done, whether they are abundant is still non-trivial 
because we have to consider the effect of the wall.}
 % against the breaking of $(\mathbb{Z}_2)_\mr{C}$.
Consequently, it is unlikely that the monopoles without the $\mathbb{Z}_2$ symmetry dominate the energy density of the universe.

We here comment on a relation of the Nambu monopole and the sphaleron in 2HDMs.
As we have studied above, the monopole is pulled by the heavier $Z$ string.
If we twist the monopole relatively to the anti-monopole, 
there could arise a repulsive force between them.
Thus, if the repulsive force and the tension of the string are balanced, 
the configuration would be a static and unstable solution of the EOM,
which is a new type of the sphaleron in the 2HDM.
This is infinitely long and has an infrared-divergent energy
while the ordinary sphaleron in 2HDMs studied in Refs.~\cite{Grant:2001at,Grant:1998ci,Moreno:1996zm,Kastening:1991nw}
is compact and has a finite energy.
It is interesting to consider if the new sphaleron can contribute to the baryon asymmetry in the universe.

%%%%%%%%%%   ACKNOWLEDGMENTS   %%%%%%%%%%

\section*{Acknowledgements} 
Y.H. would like to thank Hidefumi Matsuda for useful discussions.
% on numerical simulations.
%
This work is supported by 
the Ministry of Education, Culture, Sports, Science (MEXT)-Supported Program for the Strategic Research Foundation at Private Universities ``Topological Science (Grant No. S1511006)''. 
The work is also supported in part by JSPS Grant-in-Aid for Scientific Research 
(KAKENHI Grant No. JP16H03984 (M.~E. and M.~N.),
 No. JP19K03839 (M.~E.),
 No. JP18J22733 (Y.~H.),
 No. JP18K03655 (M.~K.),
 No. JP18H01217 (M.~N.)), 
and also by MEXT KAKENHI Grant-in-Aid for Scientific Research on Innovative Areas Topological Materials Science, No. JP15H05855 (M.~N.) 
and Discrete Geometric Analysis for Materials Design, No. JP17H06462 (M.~E.)
 from the MEXT of Japan.

%%%%%%%%%%   APPENDIX   %%%%%%%%%%
\appendix

\section{Gauge fields induced by Higgs field}
\label{165251_19Nov19}
We here derive the expression \eqref{223014_17Mar20} by minimizing Eq.~\eqref{223311_17Mar20}.
The minimization condition is given by
\begin{align}
 0 &= \f{\delta}{\delta W_i ^a}\sum_{f=1,2} \int d^3x ~|D_i \Phi_f|^2\label{000343_18Mar20} \\
 &= \f{ig}{2} \sum_f \left[ \Phi_f ^\dagger ~\sigma ^a D_i \Phi _f - (D_i\Phi_f) ^\dagger \sigma^a \Phi_f \right] \\
 &= \f{ig}{2} \sum_f \left[ \Phi_f ^\dagger \sigma ^a \overleftrightarrow{\partial_i}\Phi_f
 - ig W_i^a |\Phi_f|^2 -i g' Y_i \Phi_f^\dagger \sigma^a \Phi_f\right]\label{224335_17Mar20}.
\end{align}
By introducing currents
\begin{equation}
 J_{f,i}^a \equiv i ~\Phi_f ^\dagger~ \sigma^a \overleftrightarrow{\partial_i} \Phi_f,\label{224738_17Mar20}
\end{equation}
we can rewrite the above condition as
\begin{align}
g W_i^a + g' Y_i n^a =&  -\f{\sum_f J_{f,i}^a}{v_\mr{sum}^2},\label{231746_17Mar20}
\end{align}
where we have used $v_\mr{sum}^2 = v_1^2+v_2^2$ and Eqs.~\eqref{224610_17Mar20} and \eqref{234435_17Mar20}.

Let us calculate the currents Eqs.~\eqref{224738_17Mar20}.
Using Fierz identities (see Ref.~\cite{Nambu:1977ag}),
we have
\begin{align}
(\Phi_f^\dagger \Phi_f) J_{f,i}^a =&  \epsilon ^{abc} (\Phi_f^\dagger \sigma^b \Phi_f) \partial_i (\Phi_f^\dagger \sigma^c \Phi_f)
 + i(\Phi_f^\dagger \sigma^a \Phi_f) (\Phi_f^\dagger \overleftrightarrow{\partial_i}\Phi_f) 
\end{align}
\begin{equation}
\therefore J_{f,i}^a = v_f^2 \epsilon^{abc} n^b \partial_i n^c + n^a J_{f,i} ^0\label{231725_17Mar20}
\end{equation}
with
\begin{equation}
J_{f,i}^0 \equiv i (\Phi_f^\dagger \overleftrightarrow{\partial_i}\Phi_f)\label{230854_17Mar20}.
\end{equation}
Recalling the expressions of $\Phi_1$ and $\Phi_2$,
\begin{equation}
 \Phi_1^{\text{mon.}}=v_1
  \begin{pmatrix}
 e^{-i \varphi} \cos \f{\theta}{2} \\ \sin \f{\theta}{2}
  \end{pmatrix}, \quad
   \Phi_2^{\text{mon.}}=v_2
  \begin{pmatrix}
 \cos \f{\theta}{2} \\  e^{i \varphi}\sin \f{\theta}{2}
  \end{pmatrix},
\end{equation}
 we have
\begin{equation}
 J_{1,i}^0 = 2 v_1^2 \cos^2 \f{\theta}{2} \partial_i \varphi,\quad  J_{2,i}^0 = -2 v_2^2 \sin^2 \f{\theta}{2} \partial_i \varphi,\label{231735_17Mar20}
\end{equation}
and hence
\begin{equation}
 \sum_f J_{f,i} ^0 = v_\mr{sum}^2 (\cos \theta + \cos 2\beta) \partial_i \varphi.\label{231938_17Mar20}
\end{equation}
Substituting Eqs.~\eqref{231938_17Mar20} and \eqref{231725_17Mar20} into \eqref{231746_17Mar20},
we obtain Eq.~\eqref{223014_17Mar20}.

Note that the minimization condition associated with $Y_i$,
\begin{equation}
 0 = \f{\delta}{\delta Y_i}\sum_{f=1,2} \int d^3x ~|D_i \Phi_f|^2 ,\label{235101_17Mar20}
\end{equation}
is not independent of Eq.~\eqref{000343_18Mar20}.
To see this, we rewrite \eqref{235101_17Mar20} as
\begin{align}
 0=& \f{ig'}{2} \sum_f \left[ \Phi_f ^\dagger D_i \Phi _f - (D_i\Phi_f) ^\dagger  \Phi_f \right] \\
 =&\f{g'}{2} \sum_f \left[ i \Phi_f ^\dagger \overleftrightarrow{\partial_i} \Phi _f
 + g W_i^a \Phi_f^\dagger \sigma^a \Phi_f
 + g' Y_i \Phi_f^\dagger \Phi_f
 \right] .\label{001410_18Mar20}
\end{align}
We decompose $W_i^a$ as
\begin{equation}
 W_i^a = W_{i,\parallel}^a + W_{i,\perp}^a,\label{001418_18Mar20}
\end{equation}
where $W_{i,\parallel}^a \propto n^a$ and $ n^a W_{i,\perp}^a=0$.
From Eq.~\eqref{001410_18Mar20} and \eqref{001418_18Mar20},
we obtain a condition for $W_{i,\parallel}^a$,
which can be obtained from Eq.~\eqref{231746_17Mar20} by projecting with $n^a$.

\bibliographystyle{jhep}
\bibliography{./references}

\end{document}